\documentclass[aps,prd,twocolumn,groupedaddress,superscriptaddress,floatfix,showkeys,showpacs]{revtex4-1}

\usepackage{epsfig}
\usepackage{multirow}
\usepackage{amsmath, amssymb,mathtools}
\usepackage{color}
\usepackage{graphicx}
\usepackage{dcolumn}
\usepackage{bm}
\usepackage{subfigure}
\usepackage[utf8]{inputenc}
\usepackage[T1]{fontenc}
\usepackage{tikz}
\usepackage{soul}
\usepackage{appendix}

\graphicspath{{Figures/}}

\def\bC{{\beta_{_{C}}}}

\begin{document}

\title{Josephson junction-based axion detection through resonant activation}

\author{Roberto Grimaudo \thanks{e-mail: roberto.grimaudo01@unipa.it}}
\affiliation{Dipartimento di Fisica e Chimica ``E. Segr\`e'', Universit\`a degli Studi di Palermo, Viale delle Scienze Ed. 18, I-90128 Palermo, Italy}

\author{Davide Valenti \thanks{e-mail: davide.valenti@unipa.it}}
\affiliation{Dipartimento di Fisica e Chimica ``E. Segr\`e'', Universit\`a degli Studi di Palermo, Viale delle Scienze Ed. 18, I-90128 Palermo, Italy}

\author{Bernardo Spagnolo \thanks{e-mail: bernardo.spagnolo@unipa.it}}
\affiliation{Dipartimento di Fisica e Chimica ``E. Segr\`e'', Universit\`a degli Studi di Palermo, Viale delle Scienze Ed. 18, I-90128 Palermo, Italy}
\affiliation{Lobachevskii University of Nizhnii Novgorod, 23 Gagarin Ave. Nizhnii Novgorod 603950 Russia}

\author{Giovanni Filatrella \thanks{e-mail: giovanni.filatrella@unisannio.it}}
\affiliation{Dep. of Sciences and Technologies, University of Sannio, Via De Sanctis, Benevento I-82100, Italy}
\affiliation{INFN, Sezione di Napoli Gruppo Collegato di Salerno, Complesso Universitario di Monte S. Angelo, I-80126 Napoli, Italy}

\author{Claudio Guarcello\thanks{e-mail: cguarcello@unisa.it}}
\affiliation{Dipartimento di Fisica ``E.R. Caianiello'', Universit\`a di Salerno, Via Giovanni Paolo II, 132, I-84084 Fisciano (SA), Italy}
\affiliation{INFN, Sezione di Napoli Gruppo Collegato di Salerno, Complesso Universitario di Monte S. Angelo, I-80126 Napoli, Italy}

\date{\today}

\begin{abstract}

{\color{black}We discuss the resonant activation phenomenon on a Josephson junction due to the coupling of the Josephson system with axions.
We show how such an effect can be exploited for axion detection.}
A nonmonotonic behavior, with a minimum, of the mean switching time from the superconducting to the resistive state versus the ratio of the axion energy and the Josephson plasma energy is found. 
We demonstrate how variations in switching times make it possible to detect the presence of the axion field.
An experimental protocol for observing axions through their coupling with a Josephson system is proposed.

\end{abstract}

\maketitle

\section{Introduction}

{\color{black}Very recently the dark-matter axion detection has become a promising and fruitful research field \cite{Nagano21,Berlin21,Alesini21,Wang21,Chaudhuri21,Backes21,Battye20,Arvanitaki20,Braine20,Buschmann20,Nagano19, Malnou19,Du18,Brubaker17}.
Josephson systems are recognized of paramount importance as a sensitive experimental tool, as a playground for many theoretical models, and for their applications in fast, low-noise electronics~\cite{Bar82,Dev84,Dev13,Gua15,Nog16,Taf19,Ira18,Bra19,Kja20,Lee20,Wal21,Ret21,Gua17}.}
In the last years, a Josephson junction 
(JJ) has been supposed to interact with axions, the hypothetical elementary particles proposed as a possible component of cold dark matter~{\color{black}\cite{Dixit21,Murayama07,Bradley03}}, by exploiting 
the matching between the energies of the axion and the JJ~\cite{Bec13,Bec17,Yan20}. Very recently, hard X-ray emission {\color{black}from neutron stars} has been explained by axion 
emission~\cite{Des20,Bus21}. The axion's mass estimation is compatible with the values postulated by the Peccei-Quinn theory introduced in 1977 to solve 
the strong CP problem in quantum chromodynamics~\cite{Pec77}. The theory introduces a new scalar field which spontaneously breaks the symmetry at 
low energies, giving rise to an axion that suppresses the CP violation~\cite{Co20,Cha20}. Moreover, unexplained events in Josephson-based experiments~\cite{Hof04,Bae08,He11,Gol12,Bre13} 
can be well justified on the basis of the axion-JJ coupling. This hypothesis has thus paved the way to think of JJs as possible axion-detectors. However, up 
to now, no systematic investigations of resonance experimental conditions, suitable for direct Josephson-based axion detection, have been carried out. \\
\indent Here, we consider a Josephson-based detector to exploit the measurable voltage drop that appears across the device when the combined 
action of bias current and thermal fluctuations induces the switch from the superconducting to the resistive state~\cite{Pie21,Gua21,Gua17,Gua19}. In the presence of 
axion coupling, the analysis of the mean switching times (MST), $\tau_{MST}$, for the JJ reveals the occurrence of a resonance effect.
This is the axion-induced resonant activation phenomenon
characterized by a nonmonotonic behavior of $\tau_{MST}$, with a minimum, versus the ratio of the axion to the Josephson plasma energy.
{\color{black}Furthermore,} our work allows the identification of the suitable experimental conditions for a Josephson system to effectively detect such an axion-JJ resonance.
Based on these findings, an experimental procedure for observing axions coupled to a JJ system is proposed.

{\color{black}
The paper is organized as follows.
The physical characteristics and the mathematical formalism of the two subsystems, JJ and axion, and the composed axion-JJ system are presented in Secs. \ref{Sec JJ}, \ref{Sec Ax} and \ref{Sec Ax-JJ}, respectively.
In Sec. \ref{Sec Res Act}, the axion-induced resonant activation phenomenon is discussed in detail, while in Sec. \ref{Sec Exp} the outlines of two possible experimental schemes are proposed.
Finally, conclusive remarks are reported in Sec. \ref{Conc}.
}

\section{RCSJ Model} \label{Sec JJ}

We consider a superconductor-normal metal-superconductor JJ (see Appendix \ref{App RCSJ}), Fig.~\ref{fig: Device}(a), 
whose phase dynamics can be described within the resistively and capacitively shunted junction (RCSJ) model~\cite{Bar82,Gua17,Gua20,GuaBer21} as
\begin{equation}\label{RCSJnormOc}
\bC\frac{d^2 \varphi (\tau_c)}{d\tau_c^2}+ \frac{d \varphi (\tau_c)}{d\tau_c} + \sin \left [ \varphi\left ( \tau_c \right ) \right ] = i_{n}(\tau_c) + i_b.
\end{equation}
The time is normalized to the inverse of the characteristic frequency, that is $\tau_c = \omega_c~t$ with $\omega_c=\left ( 2e/\hbar \right )I_cR$.
$I_c$ is the maximum Josephson current that can flow through the device, while $i_b= I_b / I_c$ and $i_n= I_n / I_c$ are, respectively, the normalized external bias current and thermal noise current.
$\bC=\omega_c RC$ is the Stewart-McCumber parameter, with $R$ and $C$ being the normal-state resistance and 
capacitance of the JJ, respectively. A JJ can be effectively described in terms of a particle moving along a washboard potential tilted by $i_b$ (see Appendix \ref{App RCSJ}), 
see Fig.~\ref{fig: Device}(b). Increasing $i_b$, the slope of the washboard potential increases and the height of the confining potential barrier reduces, up to 
vanish altogether for $i_b=1$. Overdamped and underdamped JJs are characterized by $\bC\ll 1$ and $\bC\gg 1$, respectively.\\
\indent In this work, the random current is modeled as a standard Gaussian white noise associated to the JJ resistance, with the usual statistical properties 
$< i_n (\tau)>\, = 0$ and $< i_n (\tau) i_n (\tau+\tilde{\tau})> = 2D\,\delta (\tilde{\tau})$. The amplitude of the normalized correlation is connected with the 
physical temperature $T$ through the relation~\cite{Bar82}
\begin{eqnarray}
\label{WNAmp}
D= \frac{k_BT}{R}\frac{\omega_c}{I^2_c}.
\end{eqnarray}

\section{Axion field} \label{Sec Ax}

An axion field $a$ is characterized by two parameters, the axion misalignment angle $\theta$ and the axion coupling constant $f_a$, namely 
$a=f_a\,\theta$~\cite{Sik83}. Within the Robertson-Walker metric, which is appropriate to describe the early universe, the homogeneous equation of motion of 
the axion misalignment angle $\theta$ reads~\cite{Co20}
\begin{equation}
\frac{d^2 \theta (t)}{dt^2}+ 3H \frac{d \theta (t)}{dt} + \frac{m_a^2c^4}{\hbar^2} \sin \left [ \theta \left ( t \right ) \right ] = 0.
\label{AxionEq}
\end{equation}
Here, $H \approx 2 \times 10^{-18} ~ s^{-1}$ is the Hubble parameter and $m_a$ denotes the axion 
mass.
\\
\indent It is evident the similarity between the equations of motion governing the axion and the RCSJ systems: the axion dynamics is analogous to that of a RCSJ 
with no externally applied bias current. Besides the formal mathematical analogy between the two systems, it is physically remarkable that the parameters 
characterizing the two equations are quite similar as their order of magnitude is concerned (see Appendix \ref{App Axion}).
\begin{figure}[t!!]
\centering
{\includegraphics[width=0.4\textwidth]{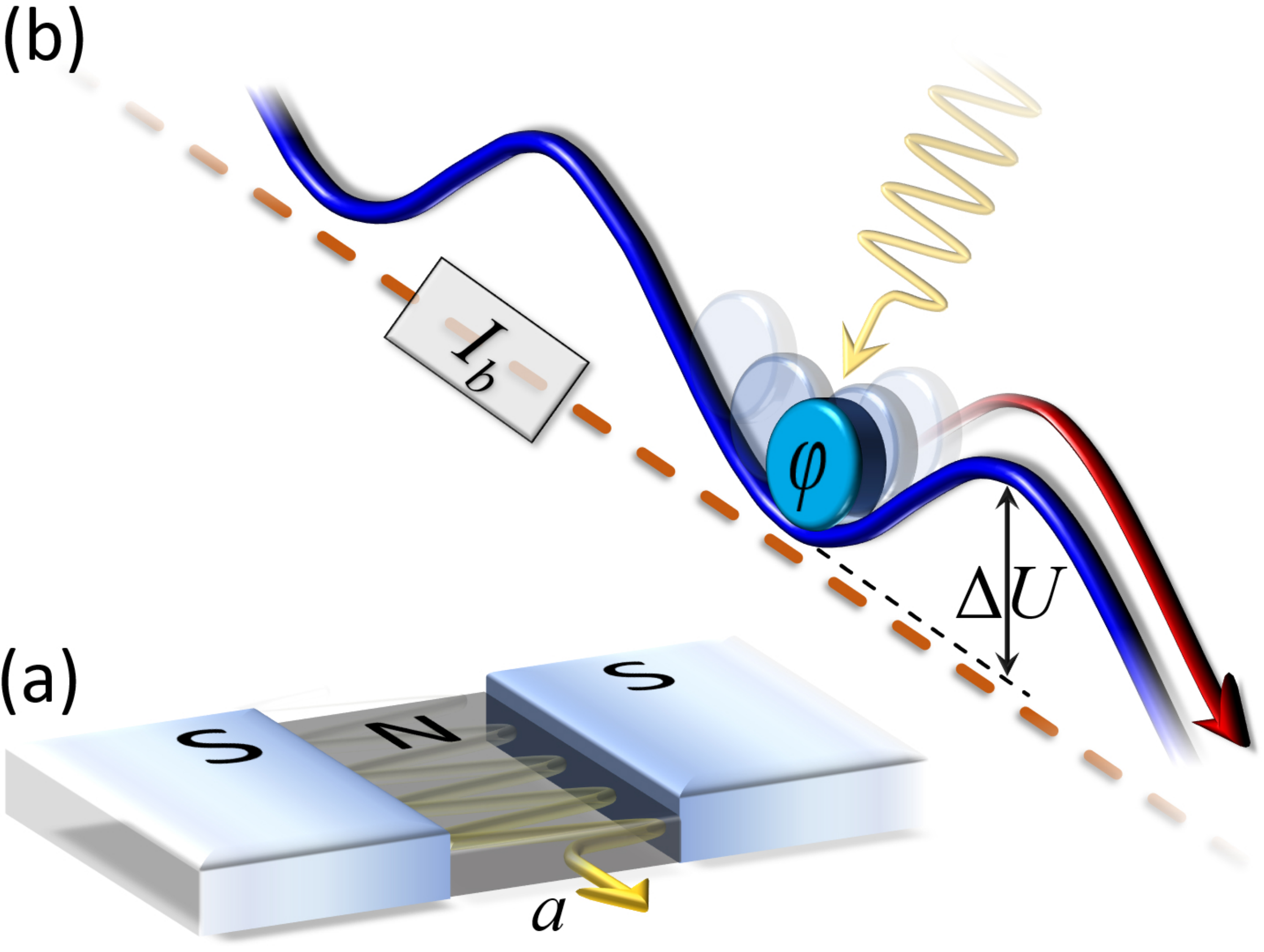}}
\caption{(a) Schematic illustration of the device. An axion field $a$ through the weak link is also represented. (b) Phase particle in a minimum of the washboard 
potential $U$, tilted by a bias current. The phase can overcome the potential barrier, $\Delta U$, rolling down along the potential because of the combined effect 
of thermal noise and axion-JJ coupling.}
\label{fig: Device}
\end{figure}

\section{Axion-JJ System}  \label{Sec Ax-JJ}

The interaction between axion and JJ can be formally written as 
\begin{subequations}\label{Orig Diff Eqs Syst}
\begin{align}
\ddot{\varphi} + a_1 \dot{\varphi} + b_1 \sin(\varphi) &= \gamma (\ddot{\theta} - \ddot{\varphi}) \label{Orig Diff Eqs Syst a},\\
\ddot{\theta} + a_2 \dot{\theta} + b_2 \sin(\theta) &= \gamma (\ddot{\varphi} - \ddot{\theta}),
\end{align}\label{Orig Diff Eqs Syst}
\end{subequations}
where $(a_1, a_2)$ and $(b_1, b_2)$ are the dissipation and frequency parameters, respectively; $\gamma$ is the coupling constant between axion and JJ and its value can be inferred from experimental quantities{\color{black}~\cite{Yan20}}.
{\color{black}
In analogy to what happens in resonant cavities, the axion-JJ coupling is supposed to be responsible for the decay of the axion into two photons, one of which (that characterized by a vanishing moment) generates electron-hole pairs which in turn create a supercurrent \cite{Dixit21,Murayama07,Bradley03,Bec13}.

By considering the presence of both a bias current and thermal fluctuations in the JJ equation, the axion-JJ system [Eqs. \eqref{Orig Diff Eqs Syst}] can be conveniently rewritten as (see Appendix \ref{App Linear})
}
\begin{subequations}\label{Diff Eqs Syst omegac}
\begin{align}
{\beta_c \over k_2}~\ddot{\varphi}+\dot{\varphi}+\sin(\varphi)+{k_1 \over k_2}~\varepsilon~\sin(\theta) &= i_b+i_n, \label{Diff Eqs Syst omegac phi} \\
{\beta_c \over k_1}~\ddot{\theta}+\dot{\varphi}+\sin(\varphi)+{k_2 \over k_1}~\varepsilon~\sin(\theta) &= i_b+i_n, \label{Diff Eqs Syst omegac theta}
\end{align}
\end{subequations}
with
\begin{equation}\label{epsilon}
k_1 = {\gamma \over 1+2\gamma}, \, k_2 = {1+\gamma \over 1+2\gamma}, \, \beta_c = \frac{\omega_c^2}{\omega_p^2}, \, 
\varepsilon = \left({m_ac^2 \over \hbar\omega_p}\right)^{\!\!\!2}\!,
\end{equation}
where $\omega_p = \sqrt{(2e I_c)/ (\hbar C)}$ is the Josephson plasma frequency.
{\color{black}
We assume that the additional sinusoidal term depending on $\theta$ in Eq. \eqref{Diff Eqs Syst omegac phi} can be ascribed to the Cooper-pair current indirectly induced by the axion entering the junction.
In other words, the axion induces an extra current term.
}
The $\varepsilon$ parameter indicates the ratio between the axion energy and the 
Josephson plasma energy, $\hbar\omega_p$, and represents our ``control knob'' to set the most convenient working point for the detection of an axion field interacting 
with the JJ. Indeed, the Josephson plasma frequency, and therefore the energy ratio $\varepsilon$, can be ``adjusted'' as needed, since $I_c$ can be lowered by raising 
the temperature~\cite{Dub01}, applying a magnetic field~\cite{Ber08} or a gate voltage~\cite{Du08,DeS19}. In this way, the system response can be tuned to achieve 
a working regime in which the switching dynamics of the axion-JJ coupled system well deviates from the Josephson response in the absence of axions. This condition 
makes the axion-JJ interaction clearly detectable.
\begin{figure}[t!!]
{\includegraphics[width=0.38\textwidth]{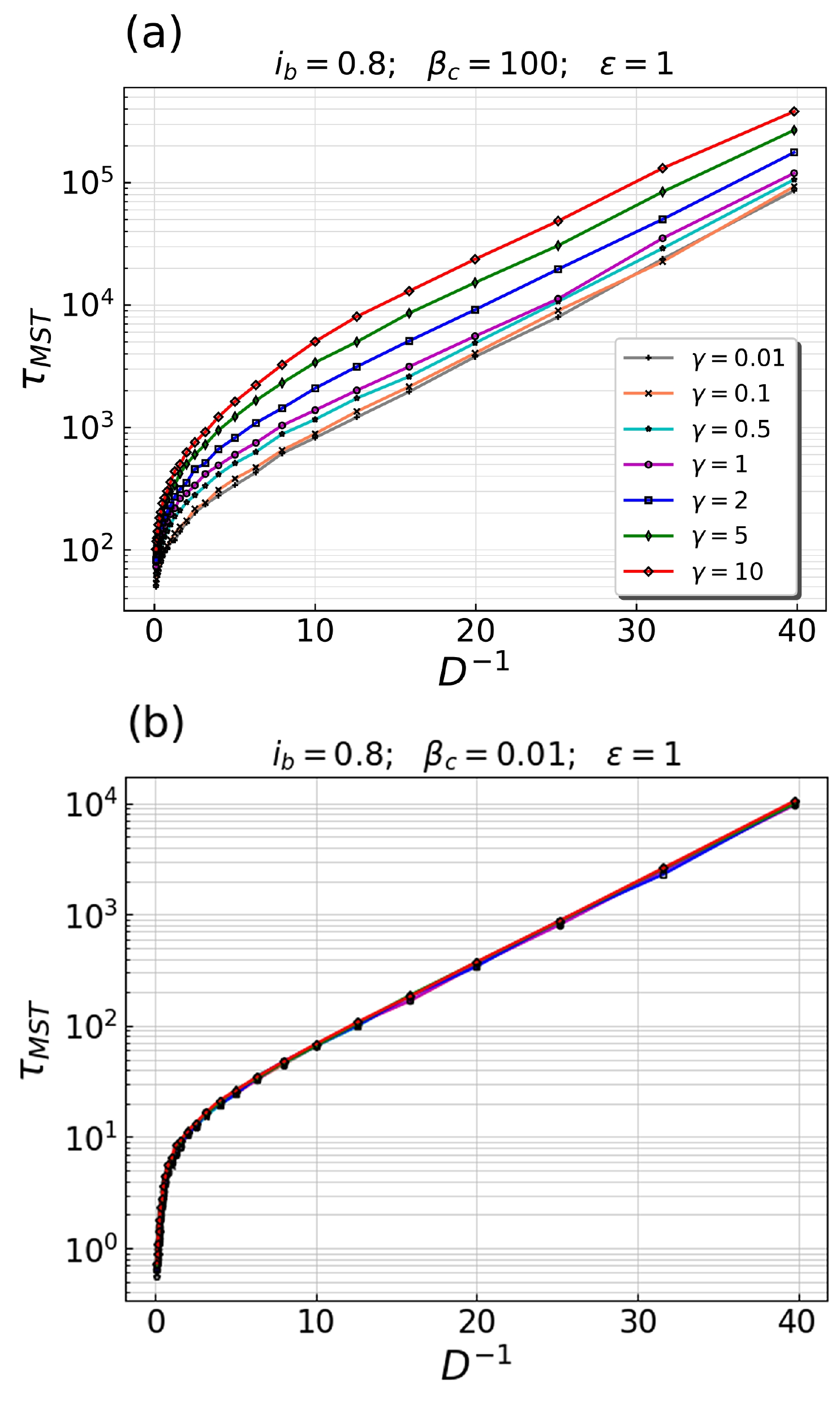}}
\caption{Semilog plot of the mean switching time ($\tau_{MST}$) versus the inverse noise intensity ($D^{-1}$), in the under- [(a)] and over- [(b)] damped regime for a JJ, 
that is {\color{black}the} coupling and decoupling regime, respectively, in the presence of thermal noise and an external bias current $i_b=0.8$, at different values of $\gamma$. The legend 
in (a) refers to both panels.} 
\label{fig: Kramers}
\end{figure}

Now we analyze how the axion affects the MST, i.e. the average time the JJ system takes to switch from the initial superconducting state 
(particle at the bottom of the well) to the resistive state. Due to random thermal fluctuations, the particle can escape from the potential well, even if the junction is 
biased by a current below the critical value (i.e., $i_b<1$). According to Kramers' theory, in the strong damping limit the average escape rate from a neighbouring 
potential barrier $\Delta \mathcal{U}(i_b)$ in the presence of a noise source, with intensity $D$ expressed by Eq.~\eqref{WNAmp}, is given in the simplest 
approximation by the following expression~\cite{Gua20}
\begin {equation}
\label{Kramersescape}
r(i_b,D)=\frac{\omega_c}{2\pi}\left ( 1-i_{b}^{2} \right )^{\frac{1}{4}}e^{-\frac{\Delta \mathcal{U}(i_b)}{D}}.
\end {equation}

In Fig.~\ref{fig: Kramers} we show the behavior of the normalized MST, $\tau_{MST}$, as a function of the inverse of the noise intensity, $D^{-1}$, 
under different damping conditions, performing $N=10^4$ independent numerical realizations from Eqs.~\eqref{Diff Eqs Syst omegac}. The linear behavior of 
$\tau_{MST}$ vs $D^ {- 1}$ characterizes a Kramers-like law. We note that if we multiply Eq.~\eqref{Orig Diff Eqs Syst a} by $\beta_c$, the latter appears only 
in the second-derivative terms, while the first-derivative term will be multiplied by the square root of $\beta_c$. Consequently, under the overdamped approximation 
($\beta_c \ll 1$) the coupling term $\beta_c\gamma(\ddot{\theta}-\ddot{\varphi})$ becomes negligible with respect to all the other terms in Eq.~\eqref{Orig Diff Eqs Syst a} 
and the two equations decouple, as it is evident by comparing the two panels in Fig.~\ref{fig: Kramers}. Therefore,
a suitable regime for the JJ system to detect axions is the underdamped regime ($\beta_c \gg 1$). \\
\indent Furthermore, the axion-coupling induces only a shift of the $\tau_{MST}$ curves upwards as the coupling parameter increases, while the slope is substantially 
unchanged (Fig.~\ref{fig: Kramers}). Therefore, only the prefactor of the Kramers-like law is influenced by $\gamma$ and not the height of the effective 
potential barrier~\cite{Graham85,Kau96}, which is given by the slope of the curves of Fig.~\ref{fig: Kramers}(a).\\

\section{Resonant activation effect} \label{Sec Res Act}

In light of the above result, it is interesting to study the dependence of the MST on the ratio $\varepsilon$ between the axion energy 
and the Josephson plasma energy. We further emphasize that the Josephson energy depends on the plasma frequency $\omega_p$ and, therefore, it can be tuned in 
experiments. However, the energy cannot be lowered at will, for example it is always necessary to ensure that the intensity of the thermal fluctuations is much lower 
than the critical current of the junction.\\
\begin{figure}[t!!]
\centering
\includegraphics[scale=0.5]{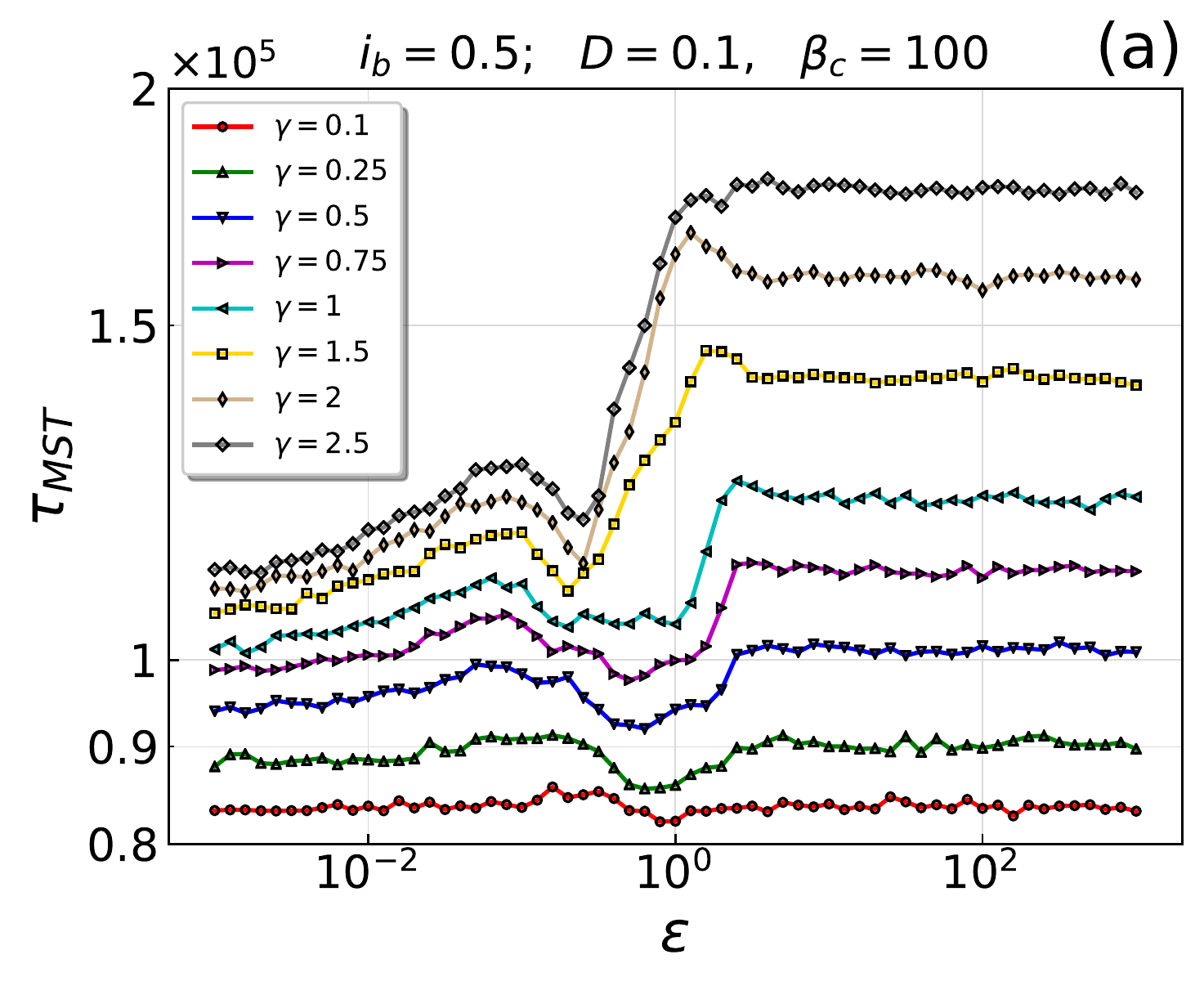} 
\\
\includegraphics[scale=0.45]{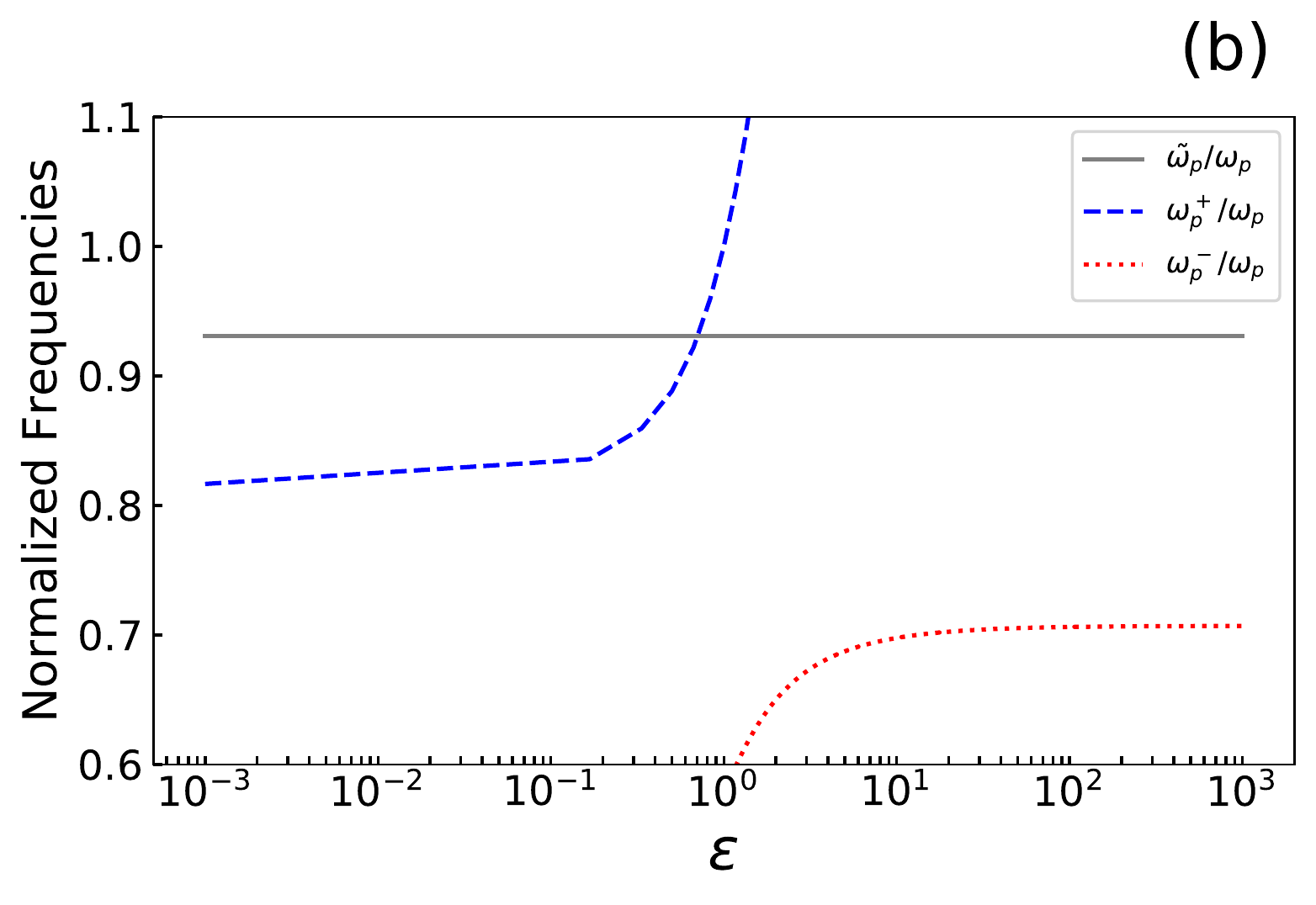} 
\caption{(a) Log-log plot of the dependence of the mean switching time ($\tau_{MST}$) on the energy ratio $\varepsilon$, in underdamped regime ($\beta_c=100$), 
with $i_b = 0.5$ and $D = 0.1$ for a JJ subject to thermal noise and coupled with an axion field.
{\color{black}The statistics is based on a set of $5 \cdot 10^4$ realizations.}
(b) Semilog plot of two normalized frequencies $\omega_p^+$ 
(dashed blue line) and $\omega_p^-$ (dotted red line)], which characterize the axion-JJ dynamics in the underdamped ($\beta_c=100$) small-oscillation 
($\dot{\varphi} \rightarrow 0$) regime with $\gamma=1.0$, \textit{vs} the ratio $\varepsilon$. The solid gray line represents the effective normalized plasma frequency 
of the system $\Tilde{\omega}_p/\omega_p=(1-i_b^2)^{1/4}$ resulting from the application of a bias current $i_b=0.5$ (see Appendix \ref{App Linear}).}
\label{fig: tau eps-matching}
\end{figure}
\indent Figure~\ref{fig: tau eps-matching}(a) clearly shows a significant nonmonotonic behavior of $\tau_{MST}$~\textit{vs}~$\varepsilon$, with a minimum in the range 
$\varepsilon \in [0.1,1]$, which is a signature of an axion-JJ resonant activation phenomenon, observed in JJs both in the absence and presence of a noise source~\cite{Dev84,Gua15,Doe92}. 
This resonant phenomenon is ascribed to the frequency-matching condition between the frequency associated to the axion angle field $\theta$ and the JJ plasma frequency (see Appendix \ref{App Linear}). In fact, the term $(k_1/k_2) \varepsilon \sin(\theta)$ in Eq.~\eqref{Diff Eqs Syst omegac phi} can be interpreted as an oscillating current for 
the Josephson system, which is responsible for the resonant activation phenomenon~\cite{Dev84,Gua15}. The minimum is less pronounced for low and high values of $\gamma$. 
In particular, for low values of the coupling parameter, $\gamma \lesssim 0.5$, the minimum is affected by the decoupling, which smoothes the curve towards a constant 
behavior; for high coupling values, $\gamma \gg 0.5$, very high values of $\tau_{MST}$, due to the confinement of the JJ phase particle ($\varepsilon \gtrsim 1$), tend to 
shallow the minimum. In the intermediate range, the minimum is more pronounced showing the resonant activation phenomenon.
{\color{black}
Indeed, by linearizing Eqs.~\eqref{Orig Diff Eqs Syst}, 
in the absence of noise, we get in the underdamped regime the expression for the frequency associated with the axion-JJ system (see Appendix \ref{App Linear})
\begin{equation}
\omega_p^+ (\gamma,\varepsilon) = \omega_p \sqrt{k_2 (\varepsilon+1) + f(\gamma,\varepsilon) \over 2 },
\end{equation}
with
\begin{equation}
f(\gamma,\varepsilon) = \sqrt{k_2^2 (\varepsilon - 1)^2 + 4 k_1^2 \varepsilon}.
\end{equation}
The frequency matching between $\omega_p^+$ and the effective plasma frequency $\Tilde{\omega}_p=\omega_p(1-i_b^2)^{1/4}$ occurs at $\varepsilon \simeq 0.7$ [see Fig.~\ref{fig: tau eps-matching}(b)],} just close to the position of the minimum in the curves of $\tau_{MST}$ \emph{vs} $\varepsilon$ in Fig.~\ref{fig: tau eps-matching}(a). The resonant matching condition 
is robust enough to be observed with a different set of parameter values (see Appendix \ref{App Linear}).
%
\indent Furthermore, the position of this local minimum depends on both the coupling parameter $\gamma$ and the applied bias current $i_b$, and moves towards lower 
values of $\varepsilon$ for higher values of $\gamma$. For low values of $i_b$ the resonant effect is more visible [Fig.~\ref{fig: tau eps-matching}(a)], while for higher bias 
values it tends to disappear as $\gamma$ increases (see Appendix \ref{App Linear}). At low noise intensity, the resonant effect is still present, and even more evident (see Appendix \ref{App Linear}).
Moreover, for $\varepsilon \ll 1$ and $\varepsilon \gg 1$, the curves approach two different plateaux. For $\varepsilon \ll 1$, the two equations describing the dynamics 
of the two systems, JJ and axion, decouple, since the effects of $\theta(t)$ on $\varphi(t)$ in Eq.~\eqref{Diff Eqs Syst omegac phi} are due to the term 
$(k_1/k_2)\varepsilon \sin(\theta)$. For $\varepsilon \gg 1$, the oscillations of $\theta(t)$ are highly damped, so as to compensate the high values of $\varepsilon$. 
This can be seen in Eq.~\eqref{Diff Eqs Syst omegac theta} where the term $(k_2/k_1) \varepsilon \sin(\theta)$ is responsible for the sinusoidal shape of the potential felt by the 
axion. For $\varepsilon \gg 1$ the potential well is extremely deep so that the axion oscillations are narrowly confined.\\
\indent Furthermore, due to the term $(k_1/k_2) \varepsilon \sin(\theta)$, which is always opposite to the bias term, the total effective current becomes smaller than $i_b$ (results 
not shown). This feature indicates that the presence of an axion tends to confine the effective phase particle representing the JJ system behavior. This explains why the value of 
the MST tends to increase for $\varepsilon \gtrsim 10^{-2}$.
Thus, the two plateaux at low and high $\varepsilon$ are somewhat different when $\gamma \gtrsim 0.5$, while for lower 
couplings the two plateaux are practically at the same level. In fact, for $\gamma \ll 1$ the weight of the term $(k_1/k_2) \varepsilon \sin(\theta)$ is lessened by the presence 
of $k_1/k_2=\gamma/(1+\gamma)$, which vanishes if $\gamma$ tends to zero and the axion-JJ equations decouple. Therefore, in the $\varepsilon \gg 1$ region a MST 
that deviates significantly from the expected unperturbed value ($\varepsilon \ll 1$) represents, together with the presence of the minimum, a \textit{hallmark of an axion-JJ interaction} and therefore of \textit{the axion detection}.
{\color{black}
The comparison between the MST measurements in the unperturbed case ($\varepsilon\ll$1), predicted also by 
Kramers theory, and those obtained for $\varepsilon \gg 1$ can lead to an estimate of $\gamma$.
In particular, first the behaviour of the MST as a function of the parameter $\epsilon$ is obtained.
Afterwords, through a comparison of the theoretical curves shown in Figs. \ref{fig: tau eps-matching}(a) and \ref{fig: tau eps-SupMat} with the experimental one, the value of the coupling parameter can be determined.
}

The observation that higher values of $\gamma$ give higher 
values of the MST, both for $\varepsilon \gg 1$ and for $\varepsilon \ll 1$, is well justified too. In fact, since $(k_1/k_2) \varepsilon \sin(\theta)$ effectively behaves as a current 
term in Eq. \eqref{Diff Eqs Syst omegac phi}, a stronger axion-JJ coupling results in an effective lower bias current which further confines the Josephson phase particle. 
This confinement, therefore, is due to both a greater axion-JJ coupling constant $\gamma$ and a greater energy ratio $\varepsilon$.\\
\indent This therefore identifies the suitable experimental conditions for a JJ-based axion detection. First, as the MST analysis is concerned, it has been shown that 
the underdamped regime is suitable to highlight the axion-induced effects on the JJ dynamics. Second, it has been found that for any value of the coupling parameter, 
according to the axion-mass estimates~\cite{Bec13,Bec17}, it is convenient to tune the plasma frequency, through the critical current, to reach the limit $\varepsilon \gg 1$. 
This allows an improved estimate of the parameter $\gamma$, thanks to the greatest spacing between the curves related to different values of the axion-JJ coupling 
for $\gamma \gtrsim 0.5$. This makes the values of $\gamma$ compatible with the experimental range of $\tau_{MST}$ more easily detectable for $\varepsilon \gg 1$. Third, and
most importantly, we have found a resonant activation phenomenon due to the frequency matching condition in the transition range $\varepsilon \approx 1$.\\ 
\indent The energy of the dark matter axion $m_a c^2$ is estimated in the energy range $\sim\!(0.006-2)\text{meV}$. To fulfill the $\varepsilon>1$ condition requires 
a suitable JJ device with a sufficiently low critical current $I_c$, which can be even further reduced by heating and/or magnetic fields. Therefore, although the value of 
$\varepsilon$ is not known precisely, it is still possible to design a setup to control its variation. This makes possible to range from the almost decoupled working 
regime ($\varepsilon\ll1$) to the well coupled one ($\varepsilon>1$). \\

\section{Possible Experimental Setup} \label{Sec Exp}

Based on the results previously shown, a technique to setup an experiment to detect the axion field is here outlined. First, as the parameter $\varepsilon$ depends on the 
JJ critical current $I_c$, Eq.~\eqref{epsilon}, by tuning $I_c$ it is possible to measure the MST deviation from the dynamical regime characterized by high $I_c$ (which entails small $\varepsilon$ and for which the axion signal is ineffective) to that characterized by low $I_c$ (which entails large $\varepsilon$ and for which the axion signal is effective).
Thus, as the critical current of the JJ is decreased, for instance by means of a magnetic field, the effects described by Eqs.~\eqref{Diff Eqs Syst omegac} 
become more and more evident [see Fig.~\ref{fig: tau eps-matching}(a)] {\it in the same experimental set-up}. Finally, by tuning the frequency matching condition to observe 
the resonant phenomenon, the axion should be revealed.

Another possible experimental setup is to consider many JJs with significantly different critical currents and to 
observe an increase in the MSTs when the critical current passes the condition $\varepsilon \approx 1$, that is, after Eq.(6), $I_c \lesssim \frac{m_a^2 c^4}{\hbar 2e/C}$.
Again, a tuning of the resonant matching condition should reveal the axion.
We note that the increase in Fig.~\ref{fig: tau eps-matching}(a) is observed in normalized units; 
the relation between the actual ($t_{MST}$) and the normalized ($\tau_{MST}$) average switching times also depends on the critical current: 
$t_{MST} = \tau_{MST}\frac{\hbar}{2eI_c R}$, according to the normalization of Eq.~\eqref{RCSJnormOc}. In other words, as $\varepsilon$ increases due to the decrease in $I_c$, the 
amplification effect on the non-normalized MSTs should be even greater than that shown in Fig.~\ref{fig: tau eps-matching}(a).\\

\section{Conclusions} \label{Conc}

We have investigated the MSTs of a JJ directly coupled to an axion field and subject to both a dc bias current and thermal fluctuations.
We have found the experimental conditions for a JJ-based axion detection: a) the underdamped regime; b) a Josephson plasma energy lower than the axion energy; c) the axion-induced resonant activation phenomenon, due to the occurrence of an effective frequency matching between axion and JJ, when the ratio of the axion energy to that of the junction falls in the range $\varepsilon\in[0.1,1]$.
Furthermore, an experimental strategy for a JJ-based axion detection is proposed.

{\color{black} 
Perhaps most importantly, we propose to reveal the axion presence through the analysis of the escape times from the superconducting initial state. 
Thus, studying the switching time statistics, we have found a resonant activation phenomenon, based on the plasma frequency, induced on the JJ by the axion that turns out to act as an effective time-dependent oscillating bias current.
}

Finally, our approach can 
be applied to different physical scenarios, like damped pendula, two capacitively coupled JJs~\cite{Blac09}, excitable coupled JJs~\cite{Hen15} and coupled qubits 
architectures for quantum computing \cite{Krantz,CirilloPrl}, paving the way to further theoretical achievements and new technological applications.

\begin{acknowledgments}
\emph{Acknowledgments.}
This work was supported by Italian Ministry of University and Research (MIUR) and the Government of the Russian Federation through Agreement No. 074-02-2018-330 (2).
\end{acknowledgments}

\appendix

\section{}

\subsection{RCSJ Model} \label{App RCSJ}

A short tunnel JJ is a quantum device formed by sandwiching a thin insulating layer between two superconducting electrodes, in which both 
lateral dimensions are smaller than the Josephson penetration depth~\cite{Bar82}. 
The dynamics of the Josephson phase $\varphi$ for a dissipative, current-biased short JJ can be studied within the RCSJ model~\cite{Bar82,Gua15,Spa17} that in non-normalized units can be written as
\begin{equation}
\left ( \frac{\Phi_0}{2\pi} \right )^{\!\!2}\!\! C \frac{d^2 \varphi}{d t^2}+\left ( \frac{\Phi_0}{2\pi} \right )^{\!\!2}\!\!\frac{1}{R} \frac{d \varphi}{d t}+\frac{d }{d \varphi}U 
= \left ( \frac{\Phi_0}{2\pi} \right ) I_N.
\label{RCSJ}
\end{equation}
Here, $U$ is the washboard potential along which the phase evolves,
\begin{equation}
U(\varphi,i_b)=E_{J_0}\left [1- \cos(\varphi) -i_b\varphi\right ],
\label{Washboard}
\end{equation}
where $E_{J_0}=\left ( \Phi_0/2\pi \right )I_c$. The resulting activation energy barrier, $\Delta U(i_b)$, 
confines the phase $\varphi$ in a metastable potential minimum and can be calculated as the difference between the maximum and minimum value of $U(\varphi,i_b)$. 
In units of $E_{J_0}$, it can be expressed as
\begin{equation}
\Delta \mathcal{U}(i_b)=\frac{\Delta U(i_b)}{E_{J_0}}=2\left [ \sqrt{1-i_b^2} -i_b\arccos(i_b)\right ].
\label{activationenergybarrier}
\end{equation}
In the phase particle picture, the term $i_b$ represents the tilting of the potential profile; increasing $i_b$ the slope of the washboard increases and the height 
$\Delta \mathcal{U}(i_b)$ of the right potential barrier reduces, until this activation energy vanishes for $i_b=1$, that is when the bias current reaches its critical value $I_c$. 

If one normalizes the time to the inverse of the characteristic frequency, that is $\tau_c = \omega_c~t$ with $\omega_c=\left ( 2e/\hbar \right )I_cR$, Eq.~\eqref{RCSJ} 
can be put in the dimensionless form
\begin{equation}
\bC\frac{d^2 \varphi (\tau_c)}{d\tau_c^2}+ \frac{d \varphi (\tau_c)}{d\tau_c} + \sin \left [ \varphi\left ( \tau_c \right ) \right ] = i_{n}(\tau_c) + i_b,
\label{RCSJnormOc}
\end{equation}
where $\bC=\omega_c RC$ is the Stewart-McCumber parameter. {Usually, the single-harmonic current-phase relation (CPR) is appropriate to describe the features of a JJ~\cite{Gol04}, i.e., the high-order harmonic terms can be neglected. However, we observe that a non-sinusoidal CPR, as in the case of a short SNS junction~\cite{Bee92}, is not expected to undermine the feasibility of the Josephson-based scheme for axion detection discussed in this work, but only to slightly affect the specific switching time values.} An overdamped junction has $\bC\ll 1$, that is a small capacitance and/or a small resistance. In contrast, a junction 
with $\bC\gg 1$ has a large capacitance and/or a large resistance, and is underdamped. Another way to obtain a dimensionless form of Eq.~\eqref{RCSJ} 
consists in normalizing with respect to the plasma frequency $\omega_p=\sqrt{2eI_c/\hbar C} \in [1,1000]~\textup{GHz}$. In this case the normalized RCSJ equation (\ref{RCSJ}) 
reads
\begin{equation}
\frac{d^2 \varphi (\tau_p)}{d\tau_p^2}+ \alpha \frac{d \varphi (\tau_p)}{d\tau_p} + \sin \left [ \varphi\left ( \tau_p \right ) \right ] = i_{n}(\tau_p) + i_b,
\label{RCSJnormOp}
\end{equation}
where $\alpha=1/(\omega_p\,R\,C)$ is the damping parameter and $\tau_p = \omega_p\,t$. With this time normalization the under- and over-damped regimes correspond 
to $\alpha \ll 1$ and $\alpha \gg 1$, respectively.

We note that normalizing with respect to the characteristic frequency $\omega_c$, as we did in our numerical simulations, the noise intensity $D$ can be simply 
expressed as the ratio of thermal energy to Josephson coupling energy $E_{J_0}$, or $D={k_BT}/{E_{J_0}}$, without any dependence on the damping. 
Normalizing instead with respect to the plasma frequency $\omega_p$, the noise intensity becomes $D=\alpha{k_BT}/{E_{J_0}}$.

In our numerical simulations, for Gaussian fluctuations of amplitude $D$, the stochastic independent increment reads
\begin{equation}
\Delta i_N \simeq \sqrt{ 2 D \Delta t\; }\; N\left(0,1 \right) .
\label{GLFincr}
\end{equation}
Here, the symbol $N\left(0, 1 \right)$ indicates a random function Gaussianly distributed with zero mean and unit standard deviation. The stochastic integration of 
Eqs.~\eqref{RCSJnormOc} or~\eqref{RCSJnormOp} is performed with a finite-difference explicit method, using a time integration step $\Delta t=10^{-2}$.

\subsection{Axion paramaters} \label{App Axion}

An axion field $a$ can be formally written as  $a=f_a\,\theta$~\cite{Sik83,Vis13}, where  $\theta$ is the axion misalignment angle and $f_a$ the axion coupling 
constant. The axion's misalignment angle dynamics obeys the following homogeneous equation of motion~\cite{Co20}
\begin{equation}
\frac{d^2 \theta (t)}{dt^2}+ {\color{black}3}H \frac{d \theta (t)}{dt} + \frac{m_a^2c^4}{\hbar^2} \sin \left [ \theta \left ( t \right ) \right ] = 0,
\label{AxionEq}
\end{equation}
where $m_a$ denotes the axion mass and $H \approx 2 \times 10^{-18} ~ s^{-1}$ the Hubble parameter. The typical ranges of parameters that are allowed for 
dark matter axions are \cite{sik09,Duf09}
\begin{equation}
3 \times 10^9 ~ \text{GeV} \leq f_a \leq 10^{12} ~ \text{GeV}.
\end{equation}
and
\begin{equation}\label{EqA09}
6 \times 10^{-6} ~ \text{eV} \leq m_a c^2 \leq 2 \times 10^{-3} ~ \text{eV}.
\end{equation}
The prediction of the axion's mass based on the average of the results obtained from five independent condensed matter experiments 
is~\cite{Hof04, Gol12, He11, Bae08, Bre13}
\begin{equation}\label{EqA10}
m_a c^2 = (106 \pm 6) \mu eV.
\end{equation}
{
These energies values permit to estimate the corresponding values of the energy ratio $\varepsilon = \left({m_ac^2 / \hbar\omega_p}\right)^2$, see Eq.~\eqref{epsilon}.
To do this, first of all the main parameters of Josephson must be set: for example, we can generically choose the values $R_a=500~\Omega\mu\textup{m}^2$, $C_s=100~f\textup{F}/\mu \textup{m}^2$, and $J_c= 10^6\textup{A}/\textup{m}^2$ for the resistance per area, the specific capacitance, and the critical current density, respectively, which gives a plasma frequency $\omega_p\simeq170~\textup{GHz}$ and a Stewart-McCumber parameter $\bC\simeq80$.}

{Then, from Eq.~\eqref{EqA09} we get $\varepsilon\in(0.003 - 300)$, which is approximately the range of values we have explored in this work. Moreover, these values match the almost decoupled working regime ($\varepsilon\ll1$) and the well coupled one ($\varepsilon>1$), at which the $\tau_{MST}$ \emph{vs} $\varepsilon$ curves approach two different plateaux. Finally, the average value in Eq.~\eqref{EqA10} corresponds to an energy ratio $\varepsilon \sim 0.85$, i.e., a value just close to the resonant matching condition discussed in Sec.~\ref{Sec Res Act}. }

\section{Linearization and Frequency Matching Condition} \label{App Linear}

\begin{figure*}[t!!]
{\includegraphics[width=0.84\textwidth]{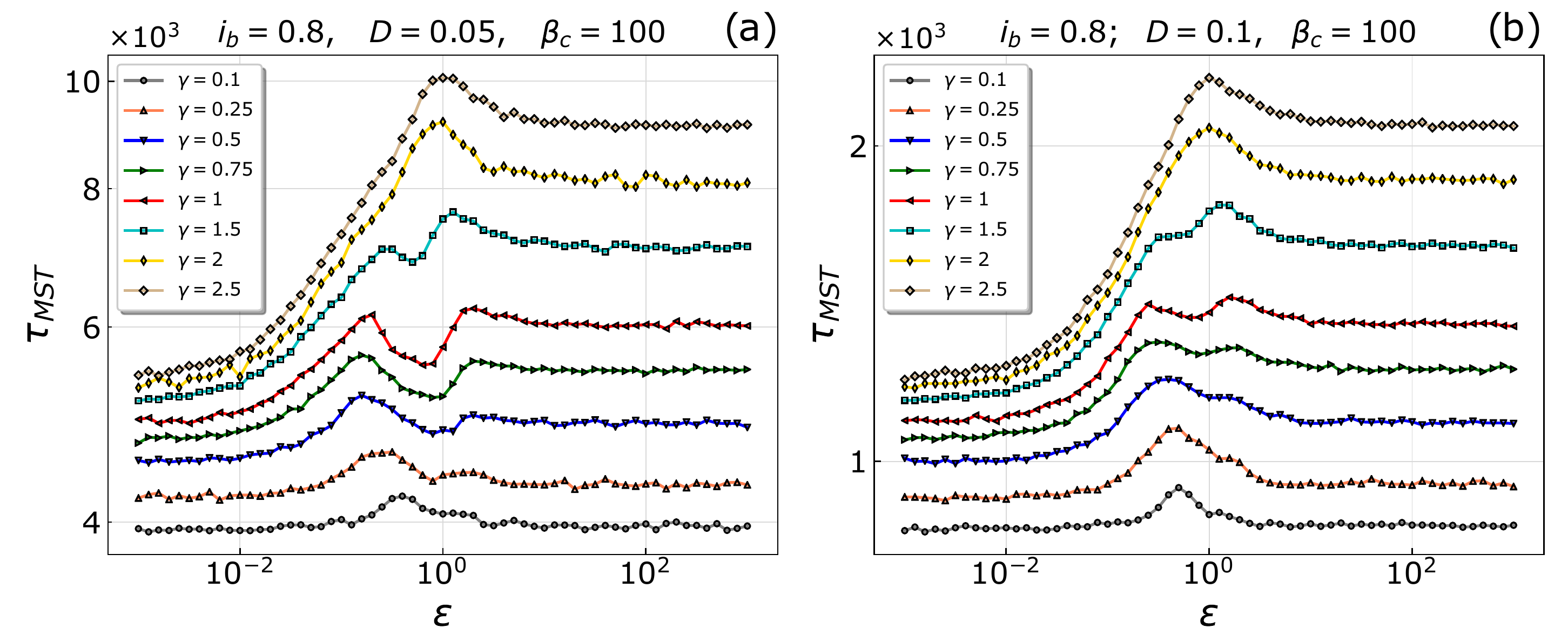}}
\caption{Log-log plot of the dependence of the mean switching time $\tau_{MST}$ on the axion-JJ coupling $\varepsilon$, in underdamped regime ($\beta_c=100$), 
with $i_b = 0.8$ and (a) $D = 0.05$, (b) $D = 0.1$ for a JJ subject to thermal noise and coupled with an axion field.
{\color{black}The statistics is based on a set of $5 \cdot 10^4$ realizations.}}
\label{fig: tau eps-SupMat}
\end{figure*}

The interaction model proposed for a coupled axion-JJ system reads [Eqs. (4) of the main text]
\begin{subequations}
\label{App Orig Diff Eqs Syst}
\begin{align}
& \ddot{\varphi} + a_1 \dot{\varphi} + b_1 \sin(\varphi) = \gamma (\ddot{\theta} - \ddot{\varphi}),\\
& \ddot{\theta} + a_2 \dot{\theta} + b_2 \sin(\theta) = \gamma (\ddot{\varphi} - \ddot{\theta}),
\end{align}
\end{subequations}
where $\gamma$ is the coupling parameter. In the presence of both a bias and a stochastic current, by normalizing with respect to the squared plasma 
frequency $\omega_p^2=b_1$, the system (\ref{Orig Diff Eqs Syst}) can be rewritten as
\begin{subequations}
\label{Diff Eqs Syst omegap}
\begin{align}
&\ddot{\varphi}+k_2~\alpha~\dot{\varphi}+k_2~\sin(\varphi)+k_1~\varepsilon~\sin(\theta)=k_2[i_b+i_n], \label{Diff Eqs Syst omegap phi}\\
&\ddot{\theta}+k_1~\alpha~\dot{\varphi}+k_1\sin(\varphi)+k_2~\varepsilon~\sin(\theta)=k_1[i_b+i_n],\label{Diff Eqs Syst omegap theta}
\end{align}
\end{subequations}
with
\begin{subequations}
\begin{align}
&\tau_p=\omega_p~t,\quad
&\alpha={a_1 \over \sqrt{b_1}}={1 \over \omega_pRC}\approx 10^0-10^1,
\end{align}
\label{parameters}
\end{subequations}
and the term proportional to $\dot{\theta}$ can be neglected, since
\begin{equation}
\begin{aligned}
{a_2 \over \sqrt{b_1}}={3H \over \omega_p}\approx 10^{-30}\sim 0.
\end{aligned}
\end{equation}
Normalizing with respect to the characteristic frequency $\omega_c$, the system of differential equations becomes 
\begin{subequations}
\label{App Diff Eqs Syst omegac}
\begin{align}
&\beta_c~\ddot{\varphi}+k_2~\dot{\varphi}+k_2~\sin(\varphi)+k_1~\varepsilon~\sin(\theta)=k_2[i_b+i_n], \\
&\beta_c~\ddot{\theta}+k_1~\dot{\varphi}+k_1\sin(\varphi)+k_2~\varepsilon~\sin(\theta)=k_1[i_b{\color{black}+}i_n],
\end{align}
\end{subequations}
with $\tau_c=\omega_c~t$. 

In this work, by numerically solving Eqs.~\eqref{Diff Eqs Syst omegac}, we report the calculation of the MST as a function of the energy ratio $\varepsilon$. Specifically, in Fig.~\ref{fig: tau eps-SupMat} the curves 
of MST versus $\varepsilon$ are shown for a higher bias value, $i_b = 0.8$, with respect to that shown in Fig.~3(a) of the main text and for two noise intensity values, 
namely $D = 0.05$ and $D=0.1$. These curves show that the resonant phenomenon tends to disappear for higher bias values and that it is still present and more evident at low 
noise intensity. Indeed, the resonant activation phenomenon is observed in the presence and in the absence of a noise source, see Refs.~\cite{Dev84,Gua15,Doe92} (and references therein).
\begin{figure}[t!!]
{\includegraphics[width=0.45\textwidth]{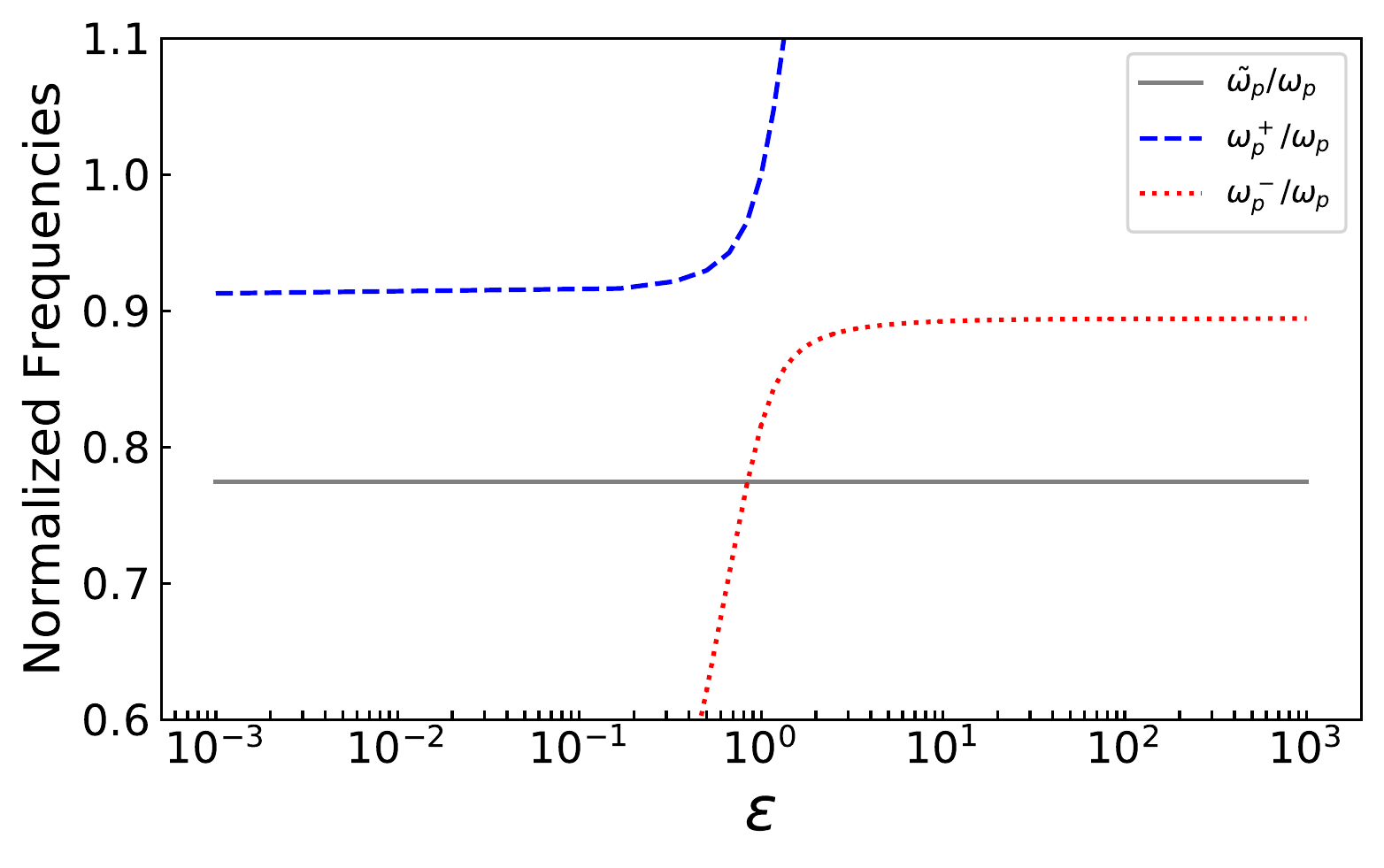}}
\caption{Dependence of the two normalized frequencies [$\omega_p^+$ (dashed blue line) and $\omega_p^-$ (dotted red line)], which characterize the axion-JJ 
dynamics in the underdamped ($\beta_c=100$) small-oscillation ($\dot{\varphi} \rightarrow 0$) regime, on the parameter $\varepsilon$, the ratio of the axion 
energy to the Josephson plasma energy, for the axion-JJ coupling $\gamma=0.25$. The solid gray line represents the effective normalized plasma frequency of 
the system $\Tilde{\omega}_p/\omega_p=(1-i_b^2)^{1/4}$ resulting from the application of a bias current $i_b=0.8$.}
\label{fig: Small-Oscillation Freqs}
\end{figure}

Let us consider the limit of small oscillations for both the JJ and the axion, in the absence of a noise source. In this case, we can approximate $\sin{\varphi} \sim \varphi$ and 
$\sin{\theta} \sim \theta$ in Eqs.~\eqref{Diff Eqs Syst omegac}. The resulting linearized system reads
\begin{subequations}
\label{Linearised System oc}
\begin{align}
\beta \ddot{\varphi} + \dot{\varphi} + \varphi + k \varepsilon \theta &= i_b, \\
\beta \ddot{\theta} + k \dot{\varphi} + k \varphi + \varepsilon \theta &= k i_b,
\end{align}
\end{subequations}
with $\beta=\beta_c/k_2$ and $k=k_1/k_2=\gamma/1+\gamma$. In the overdamped regime ($\beta_c \ll 1$), the first term can be neglected in both equations.
By adding and subtracting the two equations, we obtain
\begin{subequations}
\begin{align}
\dot{\varphi} + \varphi - i_b &= -\varepsilon \theta , \\
\dot{\varphi} + \varphi - i_b &= \varepsilon \theta,
\end{align}
\end{subequations}
which implies
\begin{subequations}
\label{Sol lin sys overd}
\begin{align}
\dot{\varphi} + \varphi - i_b=0 \qquad &\rightarrow \qquad \varphi(\tau_c) = i_b + {\varphi_0 \over 2} e^{-\tau_c}, \\
\varepsilon \theta = 0 \qquad &\rightarrow \qquad \theta(\tau_c)=0.
\end{align}
\end{subequations}
Therefore, the two equations describing the dynamics of the two systems decouple. This indicates that the overdamped linearized regime is unsuitable for axion detection.\\ 
\indent In the underdamped regime the normalization with respect to $\omega_p$ allows to more easily interpret the frequency-matching phenomenon.
{\color{black}Indeed, by putting $\alpha=0$ (underdamped regime) in Eqs. \eqref{Diff Eqs Syst omegap} and neglecting the noisy fluctuating current term, the linearized system, this time, remains coupled}
\begin{subequations}\label{Linearised System}
\begin{align}
\ddot{\varphi} + k_2 \varphi + k_1 \varepsilon \theta &= k_2 i_b, \\
\ddot{\theta} + k_1 \varphi + k_2 \varepsilon \theta &= k_1 i_b.
\end{align}
\end{subequations}
\indent The analytical solutions, for the initial conditions considered in the main text, namely $[\varphi(0),\dot{\varphi}(0),\theta(0),\dot{\theta}(0)]=[\varphi_0,0,0,0]$, are 
\begin{subequations}\label{Sol eps neq 1}
\begin{align}
\varphi(t) = i_b + (\varphi_0 - i_b) \big[ & A_-(\gamma,\varepsilon)\cos(\omega_p^+ ~ t) + \nonumber \\ 
& \varepsilon A_+(\gamma,\varepsilon)\cos(\omega_p^- ~ t) \big], \\
\theta(t) = (\varphi_0 - i_b) B(\gamma,\varepsilon) & \left[ \cos(\omega_p^+ ~ t) + \cos(\omega_p^- ~ t) \right]
\end{align}
\end{subequations}
where
\begin{subequations}
\begin{align}
\omega_p^\pm (\gamma,\varepsilon) =& \omega_p \sqrt{k_2 (\varepsilon+1) \pm f(\gamma,\varepsilon) \over 2 }, \\
f(\gamma,\varepsilon) =& \sqrt{k_2^2 (\varepsilon - 1)^2 + 4 k_1^2 \varepsilon}, \\
A_\pm (\gamma,\varepsilon) =& \frac{2K_1^2 + K_2 \pm  K_2^2 (\varepsilon-1) }{ 1 + K_2 (\varepsilon + 1)}, \\
B(\gamma,\varepsilon) = K_1 =& {k_1 \over f(\gamma,\varepsilon)}, \quad K_2 = {k_2 \over f(\gamma,\varepsilon)}.
\end{align}
\end{subequations}

In Fig.~\ref{fig: Small-Oscillation Freqs} we show the behavior of the frequencies $\omega_p^\pm (\gamma,\varepsilon)$, in units of $\omega_p$, as a function of $\varepsilon$ 
at $\gamma=0.25$ and $i_b=0.8$. The black solid line indicates the bias-dependent plasma frequency $\Tilde{\omega}_p/\omega_p=(1-i_b^2)^{1/4}$. In 
Eq.~\eqref{Diff Eqs Syst omegac}, the term $k_1~\varepsilon~\sin(\theta)$ can be considered as an \textit{oscillating drive}, with two specific characteristic frequencies given by 
$\omega_p^\pm$. Therefore, a resonant effect on the switching dynamics is expected when one of the two characteristic frequencies of the \textit{oscillating drive} and the 
Josephson plasma frequency match. This is the resonant activation phenomenon observed both in the absence and in the presence of a noise source~\cite{Dev84,Gua15}. 
In particular, in Fig.~\ref{fig: Small-Oscillation Freqs} it is shown the expected frequency matching at $\varepsilon \simeq 0.7$, that is just close to the position of the central 
minimum in the curves of $\tau_{MST}$ \emph{vs} $\varepsilon$ in Fig.2 of the main text. Here, the frequency matching is with $\omega^-$ and the parameter values are 
different from those used to get the curves of Fig. \ref{fig: tau eps-matching} in the main text. This indicates that the resonant matching condition is robust enough to be observed with a different 
set of parameter values.

For $\varepsilon=1$ we get the following simpler expressions
\begin{subequations}
\label{Sol eps eq 1}
\begin{align}
\varphi(\tau_p) =& i_b + {\varphi_0 - i_b \over 2} \left[ \cos\left( {\tau_p} \right) + \cos\left( {\tau_p \over \sqrt{1+2\gamma}} \right) \right], \\
\theta(\tau_p) =& { \varphi_0 - i_b \over 2 } \left[ \cos\left( {\tau_p} \right) - \cos\left( {\tau_p \over \sqrt{1+2\gamma}} \right) \right].
\end{align}
\end{subequations}
In this case, for small oscillations, the axion and the JJ are characterized by the same two frequencies: $\omega_1=\omega_p$ and $\omega_2=\omega_p/\sqrt{1+2\gamma}$, 
and the time evolutions of the two solutions appear very similar.

\bibliography{biblio07}

\begin{thebibliography}{66}
\expandafter\ifx\csname natexlab\endcsname\relax\def\natexlab#1{#1}\fi
\expandafter\ifx\csname bibnamefont\endcsname\relax
  \def\bibnamefont#1{#1}\fi
\expandafter\ifx\csname bibfnamefont\endcsname\relax
  \def\bibfnamefont#1{#1}\fi
\expandafter\ifx\csname citenamefont\endcsname\relax
  \def\citenamefont#1{#1}\fi
\expandafter\ifx\csname url\endcsname\relax
  \def\url#1{\texttt{#1}}\fi
\expandafter\ifx\csname urlprefix\endcsname\relax\def\urlprefix{URL }\fi
\providecommand{\bibinfo}[2]{#2}
\providecommand{\eprint}[2][]{\url{#2}}

\bibitem[{\citenamefont{Nagano et~al.}(2021)\citenamefont{Nagano, Nakatsuka,
  Morisaki, Fujita, Michimura, and Obata}}]{Nagano21}
\bibinfo{author}{\bibfnamefont{K.}~\bibnamefont{Nagano}},
  \bibinfo{author}{\bibfnamefont{H.}~\bibnamefont{Nakatsuka}},
  \bibinfo{author}{\bibfnamefont{S.}~\bibnamefont{Morisaki}},
  \bibinfo{author}{\bibfnamefont{T.}~\bibnamefont{Fujita}},
  \bibinfo{author}{\bibfnamefont{Y.}~\bibnamefont{Michimura}},
  \bibnamefont{and} \bibinfo{author}{\bibfnamefont{I.}~\bibnamefont{Obata}},
  \bibinfo{journal}{Phys. Rev. D} \textbf{\bibinfo{volume}{104}},
  \bibinfo{pages}{062008} (\bibinfo{year}{2021}).

\bibitem[{\citenamefont{Berlin et~al.}(2021)\citenamefont{Berlin, D'Agnolo,
  Ellis, and Zhou}}]{Berlin21}
\bibinfo{author}{\bibfnamefont{A.}~\bibnamefont{Berlin}},
  \bibinfo{author}{\bibfnamefont{R.~T.} \bibnamefont{D'Agnolo}},
  \bibinfo{author}{\bibfnamefont{S.~A.~R.} \bibnamefont{Ellis}},
  \bibnamefont{and} \bibinfo{author}{\bibfnamefont{K.}~\bibnamefont{Zhou}},
  \bibinfo{journal}{Phys. Rev. D} \textbf{\bibinfo{volume}{104}},
  \bibinfo{pages}{L111701} (\bibinfo{year}{2021}).

\bibitem[{\citenamefont{Alesini et~al.}(2021)\citenamefont{Alesini, Braggio,
  Carugno, Crescini, D'Agostino, Di~Gioacchino, Di~Vora, Falferi, Gambardella,
  Gatti et~al.}}]{Alesini21}
\bibinfo{author}{\bibfnamefont{D.}~\bibnamefont{Alesini}},
  \bibinfo{author}{\bibfnamefont{C.}~\bibnamefont{Braggio}},
  \bibinfo{author}{\bibfnamefont{G.}~\bibnamefont{Carugno}},
  \bibinfo{author}{\bibfnamefont{N.}~\bibnamefont{Crescini}},
  \bibinfo{author}{\bibfnamefont{D.}~\bibnamefont{D'Agostino}},
  \bibinfo{author}{\bibfnamefont{D.}~\bibnamefont{Di~Gioacchino}},
  \bibinfo{author}{\bibfnamefont{R.}~\bibnamefont{Di~Vora}},
  \bibinfo{author}{\bibfnamefont{P.}~\bibnamefont{Falferi}},
  \bibinfo{author}{\bibfnamefont{U.}~\bibnamefont{Gambardella}},
  \bibinfo{author}{\bibfnamefont{C.}~\bibnamefont{Gatti}},
  \bibnamefont{et~al.}, \bibinfo{journal}{Phys. Rev. D}
  \textbf{\bibinfo{volume}{103}}, \bibinfo{pages}{102004}
  (\bibinfo{year}{2021}).

\bibitem[{\citenamefont{Wang et~al.}(2021)\citenamefont{Wang, Bi, and
  Yin}}]{Wang21}
\bibinfo{author}{\bibfnamefont{J.-W.} \bibnamefont{Wang}},
  \bibinfo{author}{\bibfnamefont{X.-J.} \bibnamefont{Bi}}, \bibnamefont{and}
  \bibinfo{author}{\bibfnamefont{P.-F.} \bibnamefont{Yin}},
  \bibinfo{journal}{Phys. Rev. D} \textbf{\bibinfo{volume}{104}},
  \bibinfo{pages}{103015} (\bibinfo{year}{2021}).

\bibitem[{\citenamefont{Chaudhuri}(2021)}]{Chaudhuri21}
\bibinfo{author}{\bibfnamefont{S.}~\bibnamefont{Chaudhuri}},
  \bibinfo{journal}{Journal of Cosmology and Astroparticle Physics}
  \textbf{\bibinfo{volume}{2021}}, \bibinfo{pages}{033} (\bibinfo{year}{2021}).

\bibitem[{\citenamefont{Backes et~al.}(2021)\citenamefont{Backes, Palken,
  Kenany, Brubaker, Cahn, Droster, Hilton, Ghosh, Jackson, Lamoreaux
  et~al.}}]{Backes21}
\bibinfo{author}{\bibfnamefont{K.~M.} \bibnamefont{Backes}},
  \bibinfo{author}{\bibfnamefont{D.~A.} \bibnamefont{Palken}},
  \bibinfo{author}{\bibfnamefont{S.~A.} \bibnamefont{Kenany}},
  \bibinfo{author}{\bibfnamefont{B.~M.} \bibnamefont{Brubaker}},
  \bibinfo{author}{\bibfnamefont{S.~B.} \bibnamefont{Cahn}},
  \bibinfo{author}{\bibfnamefont{A.}~\bibnamefont{Droster}},
  \bibinfo{author}{\bibfnamefont{G.~C.} \bibnamefont{Hilton}},
  \bibinfo{author}{\bibfnamefont{S.}~\bibnamefont{Ghosh}},
  \bibinfo{author}{\bibfnamefont{H.}~\bibnamefont{Jackson}},
  \bibinfo{author}{\bibfnamefont{S.~K.} \bibnamefont{Lamoreaux}},
  \bibnamefont{et~al.}, \bibinfo{journal}{Nature}
  \textbf{\bibinfo{volume}{590}}, \bibinfo{pages}{238} (\bibinfo{year}{2021}).

\bibitem[{\citenamefont{Battye et~al.}(2020)\citenamefont{Battye, Garbrecht,
  McDonald, Pace, and Srinivasan}}]{Battye20}
\bibinfo{author}{\bibfnamefont{R.~A.} \bibnamefont{Battye}},
  \bibinfo{author}{\bibfnamefont{B.}~\bibnamefont{Garbrecht}},
  \bibinfo{author}{\bibfnamefont{J.~I.} \bibnamefont{McDonald}},
  \bibinfo{author}{\bibfnamefont{F.}~\bibnamefont{Pace}}, \bibnamefont{and}
  \bibinfo{author}{\bibfnamefont{S.}~\bibnamefont{Srinivasan}},
  \bibinfo{journal}{Phys. Rev. D} \textbf{\bibinfo{volume}{102}},
  \bibinfo{pages}{023504} (\bibinfo{year}{2020}).

\bibitem[{\citenamefont{Arvanitaki et~al.}(2020)\citenamefont{Arvanitaki,
  Dimopoulos, Galanis, Lehner, Thompson, and Van~Tilburg}}]{Arvanitaki20}
\bibinfo{author}{\bibfnamefont{A.}~\bibnamefont{Arvanitaki}},
  \bibinfo{author}{\bibfnamefont{S.}~\bibnamefont{Dimopoulos}},
  \bibinfo{author}{\bibfnamefont{M.}~\bibnamefont{Galanis}},
  \bibinfo{author}{\bibfnamefont{L.}~\bibnamefont{Lehner}},
  \bibinfo{author}{\bibfnamefont{J.~O.} \bibnamefont{Thompson}},
  \bibnamefont{and}
  \bibinfo{author}{\bibfnamefont{K.}~\bibnamefont{Van~Tilburg}},
  \bibinfo{journal}{Phys. Rev. D} \textbf{\bibinfo{volume}{101}},
  \bibinfo{pages}{083014} (\bibinfo{year}{2020}).

\bibitem[{\citenamefont{Braine et~al.}(2020)\citenamefont{Braine, Cervantes,
  Crisosto, Du, Kimes, Rosenberg, Rybka, Yang, Bowring, Chou
  et~al.}}]{Braine20}
\bibinfo{author}{\bibfnamefont{T.}~\bibnamefont{Braine}},
  \bibinfo{author}{\bibfnamefont{R.}~\bibnamefont{Cervantes}},
  \bibinfo{author}{\bibfnamefont{N.}~\bibnamefont{Crisosto}},
  \bibinfo{author}{\bibfnamefont{N.}~\bibnamefont{Du}},
  \bibinfo{author}{\bibfnamefont{S.}~\bibnamefont{Kimes}},
  \bibinfo{author}{\bibfnamefont{L.~J.} \bibnamefont{Rosenberg}},
  \bibinfo{author}{\bibfnamefont{G.}~\bibnamefont{Rybka}},
  \bibinfo{author}{\bibfnamefont{J.}~\bibnamefont{Yang}},
  \bibinfo{author}{\bibfnamefont{D.}~\bibnamefont{Bowring}},
  \bibinfo{author}{\bibfnamefont{A.~S.} \bibnamefont{Chou}},
  \bibnamefont{et~al.} (\bibinfo{collaboration}{ADMX Collaboration}),
  \bibinfo{journal}{Phys. Rev. Lett.} \textbf{\bibinfo{volume}{124}},
  \bibinfo{pages}{101303} (\bibinfo{year}{2020}).

\bibitem[{\citenamefont{Buschmann et~al.}(2020)\citenamefont{Buschmann, Foster,
  and Safdi}}]{Buschmann20}
\bibinfo{author}{\bibfnamefont{M.}~\bibnamefont{Buschmann}},
  \bibinfo{author}{\bibfnamefont{J.~W.} \bibnamefont{Foster}},
  \bibnamefont{and} \bibinfo{author}{\bibfnamefont{B.~R.} \bibnamefont{Safdi}},
  \bibinfo{journal}{Phys. Rev. Lett.} \textbf{\bibinfo{volume}{124}},
  \bibinfo{pages}{161103} (\bibinfo{year}{2020}).

\bibitem[{\citenamefont{Nagano et~al.}(2019)\citenamefont{Nagano, Fujita,
  Michimura, and Obata}}]{Nagano19}
\bibinfo{author}{\bibfnamefont{K.}~\bibnamefont{Nagano}},
  \bibinfo{author}{\bibfnamefont{T.}~\bibnamefont{Fujita}},
  \bibinfo{author}{\bibfnamefont{Y.}~\bibnamefont{Michimura}},
  \bibnamefont{and} \bibinfo{author}{\bibfnamefont{I.}~\bibnamefont{Obata}},
  \bibinfo{journal}{Phys. Rev. Lett.} \textbf{\bibinfo{volume}{123}},
  \bibinfo{pages}{111301} (\bibinfo{year}{2019}).

\bibitem[{\citenamefont{Malnou et~al.}(2019)\citenamefont{Malnou, Palken,
  Brubaker, Vale, Hilton, and Lehnert}}]{Malnou19}
\bibinfo{author}{\bibfnamefont{M.}~\bibnamefont{Malnou}},
  \bibinfo{author}{\bibfnamefont{D.~A.} \bibnamefont{Palken}},
  \bibinfo{author}{\bibfnamefont{B.~M.} \bibnamefont{Brubaker}},
  \bibinfo{author}{\bibfnamefont{L.~R.} \bibnamefont{Vale}},
  \bibinfo{author}{\bibfnamefont{G.~C.} \bibnamefont{Hilton}},
  \bibnamefont{and} \bibinfo{author}{\bibfnamefont{K.~W.}
  \bibnamefont{Lehnert}}, \bibinfo{journal}{Phys. Rev. X}
  \textbf{\bibinfo{volume}{9}}, \bibinfo{pages}{021023} (\bibinfo{year}{2019}).

\bibitem[{\citenamefont{Du et~al.}(2018)\citenamefont{Du, Force, Khatiwada,
  Lentz, Ottens, Rosenberg, Rybka, Carosi, Woollett, Bowring et~al.}}]{Du18}
\bibinfo{author}{\bibfnamefont{N.}~\bibnamefont{Du}},
  \bibinfo{author}{\bibfnamefont{N.}~\bibnamefont{Force}},
  \bibinfo{author}{\bibfnamefont{R.}~\bibnamefont{Khatiwada}},
  \bibinfo{author}{\bibfnamefont{E.}~\bibnamefont{Lentz}},
  \bibinfo{author}{\bibfnamefont{R.}~\bibnamefont{Ottens}},
  \bibinfo{author}{\bibfnamefont{L.~J.} \bibnamefont{Rosenberg}},
  \bibinfo{author}{\bibfnamefont{G.}~\bibnamefont{Rybka}},
  \bibinfo{author}{\bibfnamefont{G.}~\bibnamefont{Carosi}},
  \bibinfo{author}{\bibfnamefont{N.}~\bibnamefont{Woollett}},
  \bibinfo{author}{\bibfnamefont{D.}~\bibnamefont{Bowring}},
  \bibnamefont{et~al.} (\bibinfo{collaboration}{ADMX Collaboration}),
  \bibinfo{journal}{Phys. Rev. Lett.} \textbf{\bibinfo{volume}{120}},
  \bibinfo{pages}{151301} (\bibinfo{year}{2018}).

\bibitem[{\citenamefont{Brubaker et~al.}(2017)\citenamefont{Brubaker, Zhong,
  Gurevich, Cahn, Lamoreaux, Simanovskaia, Root, Lewis, Al~Kenany, Backes
  et~al.}}]{Brubaker17}
\bibinfo{author}{\bibfnamefont{B.~M.} \bibnamefont{Brubaker}},
  \bibinfo{author}{\bibfnamefont{L.}~\bibnamefont{Zhong}},
  \bibinfo{author}{\bibfnamefont{Y.~V.} \bibnamefont{Gurevich}},
  \bibinfo{author}{\bibfnamefont{S.~B.} \bibnamefont{Cahn}},
  \bibinfo{author}{\bibfnamefont{S.~K.} \bibnamefont{Lamoreaux}},
  \bibinfo{author}{\bibfnamefont{M.}~\bibnamefont{Simanovskaia}},
  \bibinfo{author}{\bibfnamefont{J.~R.} \bibnamefont{Root}},
  \bibinfo{author}{\bibfnamefont{S.~M.} \bibnamefont{Lewis}},
  \bibinfo{author}{\bibfnamefont{S.}~\bibnamefont{Al~Kenany}},
  \bibinfo{author}{\bibfnamefont{K.~M.} \bibnamefont{Backes}},
  \bibnamefont{et~al.}, \bibinfo{journal}{Phys. Rev. Lett.}
  \textbf{\bibinfo{volume}{118}}, \bibinfo{pages}{061302}
  (\bibinfo{year}{2017}).

\bibitem[{\citenamefont{Barone and Paterno}(1982)}]{Bar82}
\bibinfo{author}{\bibfnamefont{A.}~\bibnamefont{Barone}} \bibnamefont{and}
  \bibinfo{author}{\bibfnamefont{G.}~\bibnamefont{Paterno}},
  \emph{\bibinfo{title}{Physics and applications of the Josephson effect}}
  (\bibinfo{publisher}{Wiley, New York}, \bibinfo{year}{1982}).

\bibitem[{\citenamefont{Devoret et~al.}(1984)\citenamefont{Devoret, Martinis,
  Esteve, and Clarke}}]{Dev84}
\bibinfo{author}{\bibfnamefont{M.~H.} \bibnamefont{Devoret}},
  \bibinfo{author}{\bibfnamefont{J.~M.} \bibnamefont{Martinis}},
  \bibinfo{author}{\bibfnamefont{D.}~\bibnamefont{Esteve}}, \bibnamefont{and}
  \bibinfo{author}{\bibfnamefont{J.}~\bibnamefont{Clarke}},
  \bibinfo{journal}{Phys. Rev. Lett.} \textbf{\bibinfo{volume}{53}},
  \bibinfo{pages}{1260} (\bibinfo{year}{1984}).

\bibitem[{\citenamefont{Devoret and Schoelkopf}(2013)}]{Dev13}
\bibinfo{author}{\bibfnamefont{M.~H.} \bibnamefont{Devoret}} \bibnamefont{and}
  \bibinfo{author}{\bibfnamefont{R.~J.} \bibnamefont{Schoelkopf}},
  \bibinfo{journal}{Science} \textbf{\bibinfo{volume}{339}},
  \bibinfo{pages}{1169} (\bibinfo{year}{2013}).

\bibitem[{\citenamefont{Guarcello et~al.}(2015)\citenamefont{Guarcello,
  Valenti, and Spagnolo}}]{Gua15}
\bibinfo{author}{\bibfnamefont{C.}~\bibnamefont{Guarcello}},
  \bibinfo{author}{\bibfnamefont{D.}~\bibnamefont{Valenti}}, \bibnamefont{and}
  \bibinfo{author}{\bibfnamefont{B.}~\bibnamefont{Spagnolo}},
  \bibinfo{journal}{Phys. Rev. B} \textbf{\bibinfo{volume}{92}},
  \bibinfo{pages}{174519} (\bibinfo{year}{2015}).

\bibitem[{\citenamefont{Nogueira et~al.}(2016)\citenamefont{Nogueira, Nussinov,
  and van~den Brink}}]{Nog16}
\bibinfo{author}{\bibfnamefont{F.~S.} \bibnamefont{Nogueira}},
  \bibinfo{author}{\bibfnamefont{Z.}~\bibnamefont{Nussinov}}, \bibnamefont{and}
  \bibinfo{author}{\bibfnamefont{J.}~\bibnamefont{van~den Brink}},
  \bibinfo{journal}{Phys. Rev. Lett.} \textbf{\bibinfo{volume}{117}},
  \bibinfo{pages}{167002} (\bibinfo{year}{2016}).

\bibitem[{\citenamefont{Tafuri}(2019)}]{Taf19}
\bibinfo{author}{\bibfnamefont{F.}~\bibnamefont{Tafuri}},
  \emph{\bibinfo{title}{Fundamentals and Frontiers of the Josephson Effect}},
  vol. \bibinfo{volume}{286} (\bibinfo{publisher}{Springer, Cham, Switzerland},
  \bibinfo{year}{2019}).

\bibitem[{\citenamefont{Irastorza and Redondo}(2018)}]{Ira18}
\bibinfo{author}{\bibfnamefont{I.~G.} \bibnamefont{Irastorza}}
  \bibnamefont{and} \bibinfo{author}{\bibfnamefont{J.}~\bibnamefont{Redondo}},
  \bibinfo{journal}{Progress in Particle and Nuclear Physics}
  \textbf{\bibinfo{volume}{102}}, \bibinfo{pages}{89} (\bibinfo{year}{2018}).

\bibitem[{\citenamefont{Braginski}(2019)}]{Bra19}
\bibinfo{author}{\bibfnamefont{A.~I.} \bibnamefont{Braginski}},
  \bibinfo{journal}{Journal of Superconductivity and Novel Magnetism}
  \textbf{\bibinfo{volume}{32}}, \bibinfo{pages}{23} (\bibinfo{year}{2019}).

\bibitem[{\citenamefont{Kjaergaard et~al.}(2020)\citenamefont{Kjaergaard,
  Schwartz, Braumüller, Krantz, Wang, Gustavsson, and Oliver}}]{Kja20}
\bibinfo{author}{\bibfnamefont{M.}~\bibnamefont{Kjaergaard}},
  \bibinfo{author}{\bibfnamefont{M.~E.} \bibnamefont{Schwartz}},
  \bibinfo{author}{\bibfnamefont{J.}~\bibnamefont{Braumüller}},
  \bibinfo{author}{\bibfnamefont{P.}~\bibnamefont{Krantz}},
  \bibinfo{author}{\bibfnamefont{J.~I.-J.} \bibnamefont{Wang}},
  \bibinfo{author}{\bibfnamefont{S.}~\bibnamefont{Gustavsson}},
  \bibnamefont{and} \bibinfo{author}{\bibfnamefont{W.~D.}
  \bibnamefont{Oliver}}, \bibinfo{journal}{Annual Review of Condensed Matter
  Physics} \textbf{\bibinfo{volume}{11}}, \bibinfo{pages}{369}
  (\bibinfo{year}{2020}).

\bibitem[{\citenamefont{Lee et~al.}(2020)\citenamefont{Lee, Efetov, Jung,
  Ranzani, Walsh, Ohki, Taniguchi, Watanabe, Kim, Englund et~al.}}]{Lee20}
\bibinfo{author}{\bibfnamefont{G.-H.} \bibnamefont{Lee}},
  \bibinfo{author}{\bibfnamefont{D.~K.} \bibnamefont{Efetov}},
  \bibinfo{author}{\bibfnamefont{W.}~\bibnamefont{Jung}},
  \bibinfo{author}{\bibfnamefont{L.}~\bibnamefont{Ranzani}},
  \bibinfo{author}{\bibfnamefont{E.~D.} \bibnamefont{Walsh}},
  \bibinfo{author}{\bibfnamefont{T.~A.} \bibnamefont{Ohki}},
  \bibinfo{author}{\bibfnamefont{T.}~\bibnamefont{Taniguchi}},
  \bibinfo{author}{\bibfnamefont{K.}~\bibnamefont{Watanabe}},
  \bibinfo{author}{\bibfnamefont{P.}~\bibnamefont{Kim}},
  \bibinfo{author}{\bibfnamefont{D.}~\bibnamefont{Englund}},
  \bibnamefont{et~al.}, \bibinfo{journal}{Nature}
  \textbf{\bibinfo{volume}{586}}, \bibinfo{pages}{42} (\bibinfo{year}{2020}).

\bibitem[{\citenamefont{Walsh et~al.}(2021)\citenamefont{Walsh, Jung, Lee,
  Efetov, Wu, Huang, Ohki, Taniguchi, Watanabe, Kim et~al.}}]{Wal21}
\bibinfo{author}{\bibfnamefont{E.~D.} \bibnamefont{Walsh}},
  \bibinfo{author}{\bibfnamefont{W.}~\bibnamefont{Jung}},
  \bibinfo{author}{\bibfnamefont{G.-H.} \bibnamefont{Lee}},
  \bibinfo{author}{\bibfnamefont{D.~K.} \bibnamefont{Efetov}},
  \bibinfo{author}{\bibfnamefont{B.-I.} \bibnamefont{Wu}},
  \bibinfo{author}{\bibfnamefont{K.-F.} \bibnamefont{Huang}},
  \bibinfo{author}{\bibfnamefont{T.~A.} \bibnamefont{Ohki}},
  \bibinfo{author}{\bibfnamefont{T.}~\bibnamefont{Taniguchi}},
  \bibinfo{author}{\bibfnamefont{K.}~\bibnamefont{Watanabe}},
  \bibinfo{author}{\bibfnamefont{P.}~\bibnamefont{Kim}}, \bibnamefont{et~al.},
  \bibinfo{journal}{Science} \textbf{\bibinfo{volume}{372}},
  \bibinfo{pages}{409} (\bibinfo{year}{2021}).

\bibitem[{\citenamefont{Rettaroli et~al.}(2021)\citenamefont{Rettaroli,
  Alesini, Babusci, Barone, Buonomo, Beretta, Castellano, Chiarello,
  Di~Gioacchino, Felici et~al.}}]{Ret21}
\bibinfo{author}{\bibfnamefont{A.}~\bibnamefont{Rettaroli}},
  \bibinfo{author}{\bibfnamefont{D.}~\bibnamefont{Alesini}},
  \bibinfo{author}{\bibfnamefont{D.}~\bibnamefont{Babusci}},
  \bibinfo{author}{\bibfnamefont{C.}~\bibnamefont{Barone}},
  \bibinfo{author}{\bibfnamefont{B.}~\bibnamefont{Buonomo}},
  \bibinfo{author}{\bibfnamefont{M.~M.} \bibnamefont{Beretta}},
  \bibinfo{author}{\bibfnamefont{G.}~\bibnamefont{Castellano}},
  \bibinfo{author}{\bibfnamefont{F.}~\bibnamefont{Chiarello}},
  \bibinfo{author}{\bibfnamefont{D.}~\bibnamefont{Di~Gioacchino}},
  \bibinfo{author}{\bibfnamefont{G.}~\bibnamefont{Felici}},
  \bibnamefont{et~al.}, \bibinfo{journal}{Instruments}
  \textbf{\bibinfo{volume}{5}}, \bibinfo{pages}{022915} (\bibinfo{year}{2021}).

\bibitem[{\citenamefont{Guarcello et~al.}(2017)\citenamefont{Guarcello,
  Valenti, Spagnolo, Pierro, and Filatrella}}]{Gua17}
\bibinfo{author}{\bibfnamefont{C.}~\bibnamefont{Guarcello}},
  \bibinfo{author}{\bibfnamefont{D.}~\bibnamefont{Valenti}},
  \bibinfo{author}{\bibfnamefont{B.}~\bibnamefont{Spagnolo}},
  \bibinfo{author}{\bibfnamefont{V.}~\bibnamefont{Pierro}}, \bibnamefont{and}
  \bibinfo{author}{\bibfnamefont{G.}~\bibnamefont{Filatrella}},
  \bibinfo{journal}{Nanotechnology} \textbf{\bibinfo{volume}{28}},
  \bibinfo{pages}{134001} (\bibinfo{year}{2017}).

\bibitem[{\citenamefont{Dixit et~al.}(2021)\citenamefont{Dixit, Chakram, He,
  Agrawal, Naik, Schuster, and Chou}}]{Dixit21}
\bibinfo{author}{\bibfnamefont{A.~V.} \bibnamefont{Dixit}},
  \bibinfo{author}{\bibfnamefont{S.}~\bibnamefont{Chakram}},
  \bibinfo{author}{\bibfnamefont{K.}~\bibnamefont{He}},
  \bibinfo{author}{\bibfnamefont{A.}~\bibnamefont{Agrawal}},
  \bibinfo{author}{\bibfnamefont{R.~K.} \bibnamefont{Naik}},
  \bibinfo{author}{\bibfnamefont{D.~I.} \bibnamefont{Schuster}},
  \bibnamefont{and} \bibinfo{author}{\bibfnamefont{A.}~\bibnamefont{Chou}},
  \bibinfo{journal}{Phys. Rev. Lett.} \textbf{\bibinfo{volume}{126}},
  \bibinfo{pages}{141302} (\bibinfo{year}{2021}).

\bibitem[{\citenamefont{Murayama}(2007)}]{Murayama07}
\bibinfo{author}{\bibfnamefont{H.}~\bibnamefont{Murayama}}, in
  \emph{\bibinfo{booktitle}{Particle Physics and Cosmology: The Fabric of
  Spacetime}}, edited by
  \bibinfo{editor}{\bibfnamefont{F.}~\bibnamefont{Bernardeau}},
  \bibinfo{editor}{\bibfnamefont{C.}~\bibnamefont{Grojean}}, \bibnamefont{and}
  \bibinfo{editor}{\bibfnamefont{J.}~\bibnamefont{Dalibard}}
  (\bibinfo{publisher}{Elsevier}, \bibinfo{year}{2007}),
  vol.~\bibinfo{volume}{86} of \emph{\bibinfo{series}{Les Houches}}, pp.
  \bibinfo{pages}{287--347}.

\bibitem[{\citenamefont{Bradley et~al.}(2003)\citenamefont{Bradley, Clarke,
  Kinion, Rosenberg, van Bibber, Matsuki, M\"uck, and Sikivie}}]{Bradley03}
\bibinfo{author}{\bibfnamefont{R.}~\bibnamefont{Bradley}},
  \bibinfo{author}{\bibfnamefont{J.}~\bibnamefont{Clarke}},
  \bibinfo{author}{\bibfnamefont{D.}~\bibnamefont{Kinion}},
  \bibinfo{author}{\bibfnamefont{L.~J.} \bibnamefont{Rosenberg}},
  \bibinfo{author}{\bibfnamefont{K.}~\bibnamefont{van Bibber}},
  \bibinfo{author}{\bibfnamefont{S.}~\bibnamefont{Matsuki}},
  \bibinfo{author}{\bibfnamefont{M.}~\bibnamefont{M\"uck}}, \bibnamefont{and}
  \bibinfo{author}{\bibfnamefont{P.}~\bibnamefont{Sikivie}},
  \bibinfo{journal}{Rev. Mod. Phys.} \textbf{\bibinfo{volume}{75}},
  \bibinfo{pages}{777} (\bibinfo{year}{2003}).

\bibitem[{\citenamefont{Beck}(2013)}]{Bec13}
\bibinfo{author}{\bibfnamefont{C.}~\bibnamefont{Beck}}, \bibinfo{journal}{Phys.
  Rev. Lett.} \textbf{\bibinfo{volume}{111}}, \bibinfo{pages}{231801}
  (\bibinfo{year}{2013}).

\bibitem[{\citenamefont{Beck}(2017)}]{Bec17}
\bibinfo{author}{\bibfnamefont{C.}~\bibnamefont{Beck}}, \bibinfo{journal}{PoS}
  \textbf{\bibinfo{volume}{EPS-HEP2017}}, \bibinfo{pages}{058}
  (\bibinfo{year}{2017}).

\bibitem[{\citenamefont{Yan and Beck}(2020)}]{Yan20}
\bibinfo{author}{\bibfnamefont{J.}~\bibnamefont{Yan}} \bibnamefont{and}
  \bibinfo{author}{\bibfnamefont{C.}~\bibnamefont{Beck}},
  \bibinfo{journal}{Physica D: Nonlinear Phenomena}
  \textbf{\bibinfo{volume}{403}}, \bibinfo{pages}{132294}
  (\bibinfo{year}{2020}).

\bibitem[{\citenamefont{Dessert et~al.}(2020)\citenamefont{Dessert, Foster, and
  Safdi}}]{Des20}
\bibinfo{author}{\bibfnamefont{C.}~\bibnamefont{Dessert}},
  \bibinfo{author}{\bibfnamefont{J.~W.} \bibnamefont{Foster}},
  \bibnamefont{and} \bibinfo{author}{\bibfnamefont{B.~R.} \bibnamefont{Safdi}},
  \bibinfo{journal}{The Astrophysical Journal} \textbf{\bibinfo{volume}{904}},
  \bibinfo{pages}{42} (\bibinfo{year}{2020}).

\bibitem[{\citenamefont{Malte~Buschmann
  et~al.}(2021)\citenamefont{Malte~Buschmann, Co, Dessert, and Safdi}}]{Bus21}
\bibinfo{author}{\bibfnamefont{M.}~\bibnamefont{Malte~Buschmann}},
  \bibinfo{author}{\bibfnamefont{R.~T.} \bibnamefont{Co}},
  \bibinfo{author}{\bibfnamefont{C.}~\bibnamefont{Dessert}}, \bibnamefont{and}
  \bibinfo{author}{\bibfnamefont{B.~R.} \bibnamefont{Safdi}},
  \bibinfo{journal}{Physical Review Letters} \textbf{\bibinfo{volume}{126}},
  \bibinfo{pages}{021102} (\bibinfo{year}{2021}).

\bibitem[{\citenamefont{Peccei and Quinn}(1977)}]{Pec77}
\bibinfo{author}{\bibfnamefont{R.~D.} \bibnamefont{Peccei}} \bibnamefont{and}
  \bibinfo{author}{\bibfnamefont{H.~R.} \bibnamefont{Quinn}},
  \bibinfo{journal}{Phys. Rev. Lett.} \textbf{\bibinfo{volume}{38}},
  \bibinfo{pages}{1440} (\bibinfo{year}{1977}).

\bibitem[{\citenamefont{Co et~al.}(2020)\citenamefont{Co, Hall, and
  Harigaya}}]{Co20}
\bibinfo{author}{\bibfnamefont{R.~T.} \bibnamefont{Co}},
  \bibinfo{author}{\bibfnamefont{L.~J.} \bibnamefont{Hall}}, \bibnamefont{and}
  \bibinfo{author}{\bibfnamefont{K.}~\bibnamefont{Harigaya}},
  \bibinfo{journal}{Phys. Rev. Lett.} \textbf{\bibinfo{volume}{124}},
  \bibinfo{pages}{251802} (\bibinfo{year}{2020}).

\bibitem[{\citenamefont{Chang and Cui}(2020)}]{Cha20}
\bibinfo{author}{\bibfnamefont{C.-F.} \bibnamefont{Chang}} \bibnamefont{and}
  \bibinfo{author}{\bibfnamefont{Y.}~\bibnamefont{Cui}},
  \bibinfo{journal}{Phys. Rev. D} \textbf{\bibinfo{volume}{102}},
  \bibinfo{pages}{015003} (\bibinfo{year}{2020}).

\bibitem[{\citenamefont{Hoffmann et~al.}(2004)\citenamefont{Hoffmann, Lefloch,
  Sanquer, and Pannetier}}]{Hof04}
\bibinfo{author}{\bibfnamefont{C.}~\bibnamefont{Hoffmann}},
  \bibinfo{author}{\bibfnamefont{F.}~\bibnamefont{Lefloch}},
  \bibinfo{author}{\bibfnamefont{M.}~\bibnamefont{Sanquer}}, \bibnamefont{and}
  \bibinfo{author}{\bibfnamefont{B.}~\bibnamefont{Pannetier}},
  \bibinfo{journal}{Phys. Rev. B} \textbf{\bibinfo{volume}{70}},
  \bibinfo{pages}{180503} (\bibinfo{year}{2004}).

\bibitem[{\citenamefont{Bae et~al.}(2008)\citenamefont{Bae, Dinsmore~III, Sahu,
  Lee, and Bezryadin}}]{Bae08}
\bibinfo{author}{\bibfnamefont{M.-H.} \bibnamefont{Bae}},
  \bibinfo{author}{\bibfnamefont{R.~C.} \bibnamefont{Dinsmore~III}},
  \bibinfo{author}{\bibfnamefont{M.}~\bibnamefont{Sahu}},
  \bibinfo{author}{\bibfnamefont{H.-J.} \bibnamefont{Lee}}, \bibnamefont{and}
  \bibinfo{author}{\bibfnamefont{A.}~\bibnamefont{Bezryadin}},
  \bibinfo{journal}{Phys. Rev. B} \textbf{\bibinfo{volume}{77}},
  \bibinfo{pages}{144501} (\bibinfo{year}{2008}).

\bibitem[{\citenamefont{He et~al.}(2011)\citenamefont{He, Wang, and
  Chan}}]{He11}
\bibinfo{author}{\bibfnamefont{L.}~\bibnamefont{He}},
  \bibinfo{author}{\bibfnamefont{J.}~\bibnamefont{Wang}}, \bibnamefont{and}
  \bibinfo{author}{\bibfnamefont{M.~H.} \bibnamefont{Chan}},
  \bibinfo{journal}{arXiv preprint arXiv:1107.0061}  (\bibinfo{year}{2011}).

\bibitem[{\citenamefont{Golikova et~al.}(2012)\citenamefont{Golikova, H\"ubler,
  Beckmann, Batov, Karminskaya, Kupriyanov, Golubov, and Ryazanov}}]{Gol12}
\bibinfo{author}{\bibfnamefont{T.~E.} \bibnamefont{Golikova}},
  \bibinfo{author}{\bibfnamefont{F.}~\bibnamefont{H\"ubler}},
  \bibinfo{author}{\bibfnamefont{D.}~\bibnamefont{Beckmann}},
  \bibinfo{author}{\bibfnamefont{I.~E.} \bibnamefont{Batov}},
  \bibinfo{author}{\bibfnamefont{T.~Y.} \bibnamefont{Karminskaya}},
  \bibinfo{author}{\bibfnamefont{M.~Y.} \bibnamefont{Kupriyanov}},
  \bibinfo{author}{\bibfnamefont{A.~A.} \bibnamefont{Golubov}},
  \bibnamefont{and} \bibinfo{author}{\bibfnamefont{V.~V.}
  \bibnamefont{Ryazanov}}, \bibinfo{journal}{Phys. Rev. B}
  \textbf{\bibinfo{volume}{86}}, \bibinfo{pages}{064416}
  (\bibinfo{year}{2012}).

\bibitem[{\citenamefont{Bretheau et~al.}(2013)\citenamefont{Bretheau, Girit,
  Pothier, Esteve, and Urbina}}]{Bre13}
\bibinfo{author}{\bibfnamefont{L.}~\bibnamefont{Bretheau}},
  \bibinfo{author}{\bibfnamefont{{\c{C}}.~{\"O}.} \bibnamefont{Girit}},
  \bibinfo{author}{\bibfnamefont{H.}~\bibnamefont{Pothier}},
  \bibinfo{author}{\bibfnamefont{D.}~\bibnamefont{Esteve}}, \bibnamefont{and}
  \bibinfo{author}{\bibfnamefont{C.}~\bibnamefont{Urbina}},
  \bibinfo{journal}{Nature} \textbf{\bibinfo{volume}{499}},
  \bibinfo{pages}{312} (\bibinfo{year}{2013}).

\bibitem[{\citenamefont{{Piedjou Komnang} et~al.}(2021)\citenamefont{{Piedjou
  Komnang}, Guarcello, Barone, Gatti, Pagano, Pierro, Rettaroli, and
  Filatrella}}]{Pie21}
\bibinfo{author}{\bibfnamefont{A.}~\bibnamefont{{Piedjou Komnang}}},
  \bibinfo{author}{\bibfnamefont{C.}~\bibnamefont{Guarcello}},
  \bibinfo{author}{\bibfnamefont{C.}~\bibnamefont{Barone}},
  \bibinfo{author}{\bibfnamefont{C.}~\bibnamefont{Gatti}},
  \bibinfo{author}{\bibfnamefont{S.}~\bibnamefont{Pagano}},
  \bibinfo{author}{\bibfnamefont{V.}~\bibnamefont{Pierro}},
  \bibinfo{author}{\bibfnamefont{A.}~\bibnamefont{Rettaroli}},
  \bibnamefont{and}
  \bibinfo{author}{\bibfnamefont{G.}~\bibnamefont{Filatrella}},
  \bibinfo{journal}{Chaos Solitons Fract} \textbf{\bibinfo{volume}{142}},
  \bibinfo{pages}{110496} (\bibinfo{year}{2021}), ISSN
  \bibinfo{issn}{0960-0779}.

\bibitem[{\citenamefont{Guarcello et~al.}(2021)\citenamefont{Guarcello,
  Piedjou~Komnang, Barone, Rettaroli, Gatti, Pagano, and Filatrella}}]{Gua21}
\bibinfo{author}{\bibfnamefont{C.}~\bibnamefont{Guarcello}},
  \bibinfo{author}{\bibfnamefont{A.~S.} \bibnamefont{Piedjou~Komnang}},
  \bibinfo{author}{\bibfnamefont{C.}~\bibnamefont{Barone}},
  \bibinfo{author}{\bibfnamefont{A.}~\bibnamefont{Rettaroli}},
  \bibinfo{author}{\bibfnamefont{C.}~\bibnamefont{Gatti}},
  \bibinfo{author}{\bibfnamefont{S.}~\bibnamefont{Pagano}}, \bibnamefont{and}
  \bibinfo{author}{\bibfnamefont{G.}~\bibnamefont{Filatrella}},
  \bibinfo{journal}{Phys. Rev. Applied} \textbf{\bibinfo{volume}{16}},
  \bibinfo{pages}{054015} (\bibinfo{year}{2021}).

\bibitem[{\citenamefont{Guarcello et~al.}(2019)\citenamefont{Guarcello,
  Valenti, Spagnolo, Pierro, and Filatrella}}]{Gua19}
\bibinfo{author}{\bibfnamefont{C.}~\bibnamefont{Guarcello}},
  \bibinfo{author}{\bibfnamefont{D.}~\bibnamefont{Valenti}},
  \bibinfo{author}{\bibfnamefont{B.}~\bibnamefont{Spagnolo}},
  \bibinfo{author}{\bibfnamefont{V.}~\bibnamefont{Pierro}}, \bibnamefont{and}
  \bibinfo{author}{\bibfnamefont{G.}~\bibnamefont{Filatrella}},
  \bibinfo{journal}{Physical Review Applied} \textbf{\bibinfo{volume}{11}},
  \bibinfo{pages}{044078} (\bibinfo{year}{2019}).

\bibitem[{\citenamefont{Guarcello et~al.}(2020)\citenamefont{Guarcello,
  Filatrella, Spagnolo, Pierro, and Valenti}}]{Gua20}
\bibinfo{author}{\bibfnamefont{C.}~\bibnamefont{Guarcello}},
  \bibinfo{author}{\bibfnamefont{G.}~\bibnamefont{Filatrella}},
  \bibinfo{author}{\bibfnamefont{B.}~\bibnamefont{Spagnolo}},
  \bibinfo{author}{\bibfnamefont{V.}~\bibnamefont{Pierro}}, \bibnamefont{and}
  \bibinfo{author}{\bibfnamefont{D.}~\bibnamefont{Valenti}},
  \bibinfo{journal}{Physical Review Research} \textbf{\bibinfo{volume}{2}},
  \bibinfo{pages}{043332} (\bibinfo{year}{2020}).

\bibitem[{\citenamefont{Guarcello and Bergeret}(2020)}]{GuaBer21}
\bibinfo{author}{\bibfnamefont{C.}~\bibnamefont{Guarcello}} \bibnamefont{and}
  \bibinfo{author}{\bibfnamefont{F.}~\bibnamefont{Bergeret}},
  \bibinfo{journal}{Phys. Rev. Applied} \textbf{\bibinfo{volume}{13}},
  \bibinfo{pages}{034012} (\bibinfo{year}{2020}).

\bibitem[{\citenamefont{Sikivie}(1983)}]{Sik83}
\bibinfo{author}{\bibfnamefont{P.}~\bibnamefont{Sikivie}},
  \bibinfo{journal}{Phys. Rev. Lett.} \textbf{\bibinfo{volume}{51}},
  \bibinfo{pages}{1415} (\bibinfo{year}{1983}).

\bibitem[{\citenamefont{Dubos et~al.}(2001)\citenamefont{Dubos, Courtois,
  Pannetier, Wilhelm, Zaikin, and Sch\"on}}]{Dub01}
\bibinfo{author}{\bibfnamefont{P.}~\bibnamefont{Dubos}},
  \bibinfo{author}{\bibfnamefont{H.}~\bibnamefont{Courtois}},
  \bibinfo{author}{\bibfnamefont{B.}~\bibnamefont{Pannetier}},
  \bibinfo{author}{\bibfnamefont{F.~K.} \bibnamefont{Wilhelm}},
  \bibinfo{author}{\bibfnamefont{A.~D.} \bibnamefont{Zaikin}},
  \bibnamefont{and} \bibinfo{author}{\bibfnamefont{G.}~\bibnamefont{Sch\"on}},
  \bibinfo{journal}{Phys. Rev. B} \textbf{\bibinfo{volume}{63}},
  \bibinfo{pages}{064502} (\bibinfo{year}{2001}).

\bibitem[{\citenamefont{Bergeret and Cuevas}(2008)}]{Ber08}
\bibinfo{author}{\bibfnamefont{F.~S.} \bibnamefont{Bergeret}} \bibnamefont{and}
  \bibinfo{author}{\bibfnamefont{J.~C.} \bibnamefont{Cuevas}},
  \bibinfo{journal}{Journal of Low Temperature Physics}
  \textbf{\bibinfo{volume}{153}}, \bibinfo{pages}{304} (\bibinfo{year}{2008}).

\bibitem[{\citenamefont{Du et~al.}(2008)\citenamefont{Du, Skachko, and
  Andrei}}]{Du08}
\bibinfo{author}{\bibfnamefont{X.}~\bibnamefont{Du}},
  \bibinfo{author}{\bibfnamefont{I.}~\bibnamefont{Skachko}}, \bibnamefont{and}
  \bibinfo{author}{\bibfnamefont{E.~Y.} \bibnamefont{Andrei}},
  \bibinfo{journal}{Phys. Rev. B} \textbf{\bibinfo{volume}{77}},
  \bibinfo{pages}{184507} (\bibinfo{year}{2008}).

\bibitem[{\citenamefont{De~Simoni et~al.}(2019)\citenamefont{De~Simoni,
  Paolucci, Puglia, and Giazotto}}]{DeS19}
\bibinfo{author}{\bibfnamefont{G.}~\bibnamefont{De~Simoni}},
  \bibinfo{author}{\bibfnamefont{F.}~\bibnamefont{Paolucci}},
  \bibinfo{author}{\bibfnamefont{C.}~\bibnamefont{Puglia}}, \bibnamefont{and}
  \bibinfo{author}{\bibfnamefont{F.}~\bibnamefont{Giazotto}},
  \bibinfo{journal}{ACS Nano} \textbf{\bibinfo{volume}{13}},
  \bibinfo{pages}{7871} (\bibinfo{year}{2019}).

\bibitem[{\citenamefont{Graham and T\'el}(1985)}]{Graham85}
\bibinfo{author}{\bibfnamefont{R.}~\bibnamefont{Graham}} \bibnamefont{and}
  \bibinfo{author}{\bibfnamefont{T.}~\bibnamefont{T\'el}},
  \bibinfo{journal}{Phys. Rev. A} \textbf{\bibinfo{volume}{31}},
  \bibinfo{pages}{1109} (\bibinfo{year}{1985}).

\bibitem[{\citenamefont{Kautz}(1996)}]{Kau96}
\bibinfo{author}{\bibfnamefont{R.~L.} \bibnamefont{Kautz}},
  \bibinfo{journal}{Reports on Progress in Physics}
  \textbf{\bibinfo{volume}{59}}, \bibinfo{pages}{935} (\bibinfo{year}{1996}).

\bibitem[{\citenamefont{Doering and Gadoua}(1992)}]{Doe92}
\bibinfo{author}{\bibfnamefont{C.~R.} \bibnamefont{Doering}} \bibnamefont{and}
  \bibinfo{author}{\bibfnamefont{J.~C.} \bibnamefont{Gadoua}},
  \bibinfo{journal}{Phys. Rev. Lett.} \textbf{\bibinfo{volume}{69}},
  \bibinfo{pages}{2318} (\bibinfo{year}{1992}).

\bibitem[{\citenamefont{Blackburn et~al.}(2009)\citenamefont{Blackburn,
  Marchese, Cirillo, and Gr{\o}nbech-Jensen}}]{Blac09}
\bibinfo{author}{\bibfnamefont{J.~A.} \bibnamefont{Blackburn}},
  \bibinfo{author}{\bibfnamefont{J.~E.} \bibnamefont{Marchese}},
  \bibinfo{author}{\bibfnamefont{M.}~\bibnamefont{Cirillo}}, \bibnamefont{and}
  \bibinfo{author}{\bibfnamefont{N.}~\bibnamefont{Gr{\o}nbech-Jensen}},
  \bibinfo{journal}{Physical Review B} \textbf{\bibinfo{volume}{79}},
  \bibinfo{pages}{054516} (\bibinfo{year}{2009}).

\bibitem[{\citenamefont{Hens et~al.}(2015)\citenamefont{Hens, Pal, and
  Dana}}]{Hen15}
\bibinfo{author}{\bibfnamefont{C.}~\bibnamefont{Hens}},
  \bibinfo{author}{\bibfnamefont{P.}~\bibnamefont{Pal}}, \bibnamefont{and}
  \bibinfo{author}{\bibfnamefont{S.~K.} \bibnamefont{Dana}},
  \bibinfo{journal}{Physical Review E} \textbf{\bibinfo{volume}{92}},
  \bibinfo{pages}{022915} (\bibinfo{year}{2015}).

\bibitem[{\citenamefont{Krantz et~al.}(2019)\citenamefont{Krantz, Kjaergaard,
  Yan, Orlando, Gustavsson, and Oliver}}]{Krantz}
\bibinfo{author}{\bibfnamefont{P.}~\bibnamefont{Krantz}},
  \bibinfo{author}{\bibfnamefont{M.}~\bibnamefont{Kjaergaard}},
  \bibinfo{author}{\bibfnamefont{F.}~\bibnamefont{Yan}},
  \bibinfo{author}{\bibfnamefont{T.~P.} \bibnamefont{Orlando}},
  \bibinfo{author}{\bibfnamefont{S.}~\bibnamefont{Gustavsson}},
  \bibnamefont{and} \bibinfo{author}{\bibfnamefont{W.~D.}
  \bibnamefont{Oliver}}, \bibinfo{journal}{Applied Physics Reviews}
  \textbf{\bibinfo{volume}{6}}, \bibinfo{pages}{021318} (\bibinfo{year}{2019}).

\bibitem[{\citenamefont{Gr\o{}nbech-Jensen
  et~al.}(2010)\citenamefont{Gr\o{}nbech-Jensen, Marchese, Cirillo, and
  Blackburn}}]{CirilloPrl}
\bibinfo{author}{\bibfnamefont{N.}~\bibnamefont{Gr\o{}nbech-Jensen}},
  \bibinfo{author}{\bibfnamefont{J.~E.} \bibnamefont{Marchese}},
  \bibinfo{author}{\bibfnamefont{M.}~\bibnamefont{Cirillo}}, \bibnamefont{and}
  \bibinfo{author}{\bibfnamefont{J.~A.} \bibnamefont{Blackburn}},
  \bibinfo{journal}{Phys. Rev. Lett.} \textbf{\bibinfo{volume}{105}},
  \bibinfo{pages}{010501} (\bibinfo{year}{2010}).

\bibitem[{\citenamefont{Spagnolo et~al.}(2017)\citenamefont{Spagnolo,
  Guarcello, Magazz\'u, Carollo, Persano~Adorno, and Valenti}}]{Spa17}
\bibinfo{author}{\bibfnamefont{B.}~\bibnamefont{Spagnolo}},
  \bibinfo{author}{\bibfnamefont{C.}~\bibnamefont{Guarcello}},
  \bibinfo{author}{\bibfnamefont{L.}~\bibnamefont{Magazz\'u}},
  \bibinfo{author}{\bibfnamefont{A.}~\bibnamefont{Carollo}},
  \bibinfo{author}{\bibfnamefont{D.}~\bibnamefont{Persano~Adorno}},
  \bibnamefont{and} \bibinfo{author}{\bibfnamefont{D.}~\bibnamefont{Valenti}},
  \bibinfo{journal}{Entropy} \textbf{\bibinfo{volume}{19}}
  (\bibinfo{year}{2017}).

\bibitem[{\citenamefont{Golubov et~al.}(2004)\citenamefont{Golubov, Kupriyanov,
  and Il'ichev}}]{Gol04}
\bibinfo{author}{\bibfnamefont{A.~A.} \bibnamefont{Golubov}},
  \bibinfo{author}{\bibfnamefont{M.~Y.} \bibnamefont{Kupriyanov}},
  \bibnamefont{and} \bibinfo{author}{\bibfnamefont{E.}~\bibnamefont{Il'ichev}},
  \bibinfo{journal}{Rev. Mod. Phys.} \textbf{\bibinfo{volume}{76}},
  \bibinfo{pages}{411} (\bibinfo{year}{2004}).

\bibitem[{\citenamefont{Beenakker}(1992)}]{Bee92}
\bibinfo{author}{\bibfnamefont{C.~W.~J.} \bibnamefont{Beenakker}}, in
  \emph{\bibinfo{booktitle}{Low-Dimensional Electronic Systems}}, edited by
  \bibinfo{editor}{\bibfnamefont{G.}~\bibnamefont{Bauer}},
  \bibinfo{editor}{\bibfnamefont{F.}~\bibnamefont{Kuchar}}, \bibnamefont{and}
  \bibinfo{editor}{\bibfnamefont{H.}~\bibnamefont{Heinrich}}
  (\bibinfo{publisher}{Springer Berlin Heidelberg}, \bibinfo{address}{Berlin,
  Heidelberg}, \bibinfo{year}{1992}), pp. \bibinfo{pages}{78--82}.

\bibitem[{\citenamefont{Visinelli}(2013)}]{Vis13}
\bibinfo{author}{\bibfnamefont{L.}~\bibnamefont{Visinelli}},
  \bibinfo{journal}{Modern Physics Letters A} \textbf{\bibinfo{volume}{28}},
  \bibinfo{pages}{1350162} (\bibinfo{year}{2013}).

\bibitem[{\citenamefont{Sikivie and Yang}(2009)}]{sik09}
\bibinfo{author}{\bibfnamefont{P.}~\bibnamefont{Sikivie}} \bibnamefont{and}
  \bibinfo{author}{\bibfnamefont{Q.}~\bibnamefont{Yang}},
  \bibinfo{journal}{Physical Review Letters} \textbf{\bibinfo{volume}{103}},
  \bibinfo{pages}{111301} (\bibinfo{year}{2009}).

\bibitem[{\citenamefont{Duffy and van Bibber}(2009)}]{Duf09}
\bibinfo{author}{\bibfnamefont{L.~D.} \bibnamefont{Duffy}} \bibnamefont{and}
  \bibinfo{author}{\bibfnamefont{K.}~\bibnamefont{van Bibber}},
  \bibinfo{journal}{New Journal of Physics} \textbf{\bibinfo{volume}{11}},
  \bibinfo{pages}{105008} (\bibinfo{year}{2009}).

\end{thebibliography}

\end{document}